\documentclass{article}
            \usepackage{natbib}
           
           \bibpunct{}{}{,}{s}{}{}
          \begin{document}
           \title{Towards a Coherent Theory of Physics and
             Mathematics}
          \author{Paul Benioff\thanks{
           Physics Division, Argonne National Lab,
           Argonne, IL 60439; e-mail: pbenioff@anl.gov}\\[7ex]
           \parbox{28em}{\small
          As an approach to a Theory of Everything a framework for developing a
          coherent theory of mathematics and physics together is described.
          The main characteristic of such a theory is discussed: the theory must
          be valid and and sufficiently strong, and it must maximally describe its
          own validity and sufficient strength. The mathematical logical definition
          of validity is used, and sufficient strength is seen to be a necessary
          and useful concept. The requirement of maximal description of its own
          validity and sufficient strength may be useful to reject candidate coherent
          theories for which the description is less than maximal. Other aspects of a
          coherent theory discussed include universal applicability, the relation
          to the anthropic principle, and possible uniqueness. It is suggested that
          the basic properties of the physical and mathematical universes are entwined
          with and emerge with a coherent theory. Support for this includes the
          indirect reality status of properties of very small or very large far away
          systems compared to  moderate sized nearby systems. Discussion of the
          necessary physical nature of language includes  physical models of language
          and a proof that the meaning content of expressions of any axiomatizable
          theory seems to be independent of the algorithmic complexity of the theory.
          G\"{o}del maps seem to be less useful for a coherent theory than for purely
          mathematical theories because all symbols and words of any language must
          have representations as states of physical systems already in the domain
          of a coherent theory.}}
          \date{}
         \maketitle

\section{Introduction}
The goal of a final theory or a Theory of Everything (TOE) is a much sought after dream
that has occupied the attention of many physicists and philosophers.  The allure of
such a goal is perhaps most cogently shown by the title of a recent book "Dreams of a
Final Theory"$^{(}$\cite{Weinberg}$^{)}$.  The large effort and development of
superstring theory in physics to unify quantum mechanics and general
relativity$^{(}$\cite{Greene}$^{)}$ also attests to the desire for such a theory.

The existence of so much effort shows that it is not known if it is even possible to
construct a TOE. However the conclusion of one study$^{(}$\cite{Casti}$^{)}$ that there
do not seem to be any  fundamental limits on scientific knowledge  at least lends
support to searching for a TOE, even if it turns out that there are limits to
scientific knowledge.

Here the emphasis is on approaching a TOE from a direction that
emphasizes the close connection between mathematics and physics.
The idea is to work towards developing a coherent theory of
mathematics and physics by integrating mathematical logical
concepts with physical concepts. Whether such a coherent theory
can be constructed, and, if so, whether it is or is not a suitable
theory of everything, is a question for the future.

The approach taken here is perhaps closest to that of Tegmark$^{(}$\cite{Tegmark}$^{)}$
in that he also emphasizes general mathematical and physical aspects of a TOE along
with mathematical logical properties. However this work differs in not using the
ensemble aspect.  Also the equivalence between mathematical and physical existence of
systems is not used here.

The plan of the paper is to first provide a background discussion
on the relation between physics and mathematics at a foundational
level. Some the problems are outlined along with a summary of the
main foundational approaches to mathematics. There is also a very
brief description of other work on the foundational aspects of the
relation between physics and mathematics.

This is followed in Section \ref{CTMP} by a description of a
general framework for a coherent theory of mathematics and
physics. Much of the discussion is centered on the main and
possibly defining requirement for a coherent theory: The theory
{\em must be valid and sufficiently strong and it must maximally
describe its own validity and sufficient strength.}

This requirement is discussed in the context of the assumption that a coherent theory
can be, in principle at least, axiomatized using the first order predicate calculus. In
this way the tools that have been extensively developed in mathematical logic for
treating first order theories, such as G\"{o}del's completeness theorem and his two
incompleteness theorems$^{(}$\cite{Shoenfield,Smullyan,Godel}$^{)}$ and other results,
can be used and brought into close contact with physics and help to support a coherent
theory.

The restriction to first order theories would seem to be a minimal restriction as it
includes all mathematics used by physics so far. For instance, Zermelo Frankel set
theory, which includes all mathematics used to date by
physics,$^{(}$\cite{Frankel,Shoenfield,BenRAN}$^{)}$ is a first order theory. If it
turns out to be necessary to consider second order theories$^{(}$\cite{Vaananen}$^{)}$,
then it is hoped that the ideas developed here could be extended to these theories.

Validity is given the usual mathematical logical definition, that
all expressions of a coherent theory that are theorems are true.
The important role of physical implementability for the definition
of validity is noted. It is also seen that even though sufficient
strength cannot be defined, it is both a necessary and a useful
concept, especially for incomplete theories such as arithmetic and
set theory.  It is expected that any coherent theory will also be
incomplete, but it should be sufficiently strong to be recognized
as a coherent theory.

The potential role of the requirement that a coherent theory
maximally describe its own validity and sufficient strength  in
restricting or limiting candidate coherent theories is noted.  In
a more speculative vein a possible use of the requirement is
suggested in which it restricts paths of theories of increasing
strength generated by iterated extensions of a theory.

Other aspects of a coherent theory that are discussed include
universal applicability, the strong anthropic principle, and the
possible uniqueness of a coherent theory. Problems in how to
exactly define universal applicability are considered. It is noted
that a coherent theory should include all mathematical systems
used by physics, physical systems as complex as intelligent
systems, and it should exclude the possibility of pointing to a
specific physical system that is not included.

Emergence of the physical and mathematical universes and a
coherent theory of mathematics and physics is discussed in Section
\ref{EPMU}.  The position taken is that the basic properties of
the physical and mathematical universes and a coherent theory are
emergent together and mutually determined and entwined. They
should not be regarded as having an a priori independent
existence. To support this point the reality status of very small
or very large far away systems is compared with that of moderate
sized systems. The indirectness of properties of the former, in
their dependence on many layers of supporting theory and
experiment, is contrasted with directly experienced properties of
moderate sized and nearby systems.

The importance of the fact that language is physical for a
coherent theory is discussed in Section \ref{LP} and the Appendix.
The fact that all languages, formal or informal, necessarily have
physical representations as states of physical systems is
emphasized by describing quantum mechanical models of language.
Included is a discussion of the importance of efficient
implementability for operations such as generating text. A
specific model of a multistate system moving along a lattice of
quantum systems is used to illustrate this requirement.

Other aspects of the physical nature of language discussed in
Section \ref{LP} include a proof that the relation between the
meaning of language expression states and their algorithmic
information content is, at best, complex and is probably
nonexistent. The section closes with a discussion of G\"{o}del
maps and the observation that they play a more limited role in
coherent theories than in any purely mathematical theory. The
reason is that physical representations of language, which must
exist, are already in the domain of a coherent theory.
\\

\section{Background}

Physics and mathematics have a somewhat contradictory relationship
in that they are both closely related and are also disconnected.
The close relation can be seen by noting the mathematical nature
of theoretical physics and the use of theory to generate
predictions as the outcomes of mathematical calculations that can
be affirmed or refuted by experiment. The validity of a physical
theory is based on many such comparisons between  theory and
experiment. Agreement constitutes support for the theory.
Disagreement between theoretical predictions and experiment erodes
support for the theory and, for crucial experiments, may result in
the theory being abandoned.

The disconnect between physics and mathematics can be seen by
noting that, from a foundational point of view, physics takes
mathematics for granted. In many ways theoretical physics treats
mathematics much like a warehouse of different consistent axiom
systems each with their set of theorems.  If a system needed by
physics has been studied, it is taken from the warehouse, existing
theorems and results are used, and, if needed, new theorems are
proved. If theoretical physics needs a system which has not been
invented, it is created as a new system.  Then the needed theorems
are proved based on the axioms of the new system.

The problem here is that physics and mathematics are considered as separate
disciplines.  The possibility that they might be part of a larger coherent theory of
mathematics and physics together is not much discussed.  For example basic aspects such
as truth, validity, consistency, and provability are described in detail in
mathematical logic which is the study of  axiom systems and their
models$^{(}$\cite{Shoenfield}$^{)}$. The possibility that how these concepts are
described or defined may affect their use in physics, and may also even influence what
is true in physics at a very basic level has not been considered. (A very preliminary
attempt to see how these concepts might be used in quantum mechanics is made
in$^{(}$\cite{BenDTVQM}$^{)}$.)

The situation in mathematics is different.  Here the problem is that most work in
mathematics and mathematical logic is purely abstract with little attention paid to
foundational aspects of physics. The facts that mathematical reasoning is carried out
by physical systems subject to physical laws, and symbols, words, and formulas in any
language, formal or not, are physical systems in different states, is, for the most
part, ignored. In some ways the various constructivist interpretations of mathematics,
ranging from extreme intuitionism$^{(}$\cite{Heyting}$^{)}$ to more moderate
views$^{(}$\cite{Bishop,Beeson}$^{)}$ (see also$^{(}$\cite{Frankel}$^{)}$) do
acknowledge this problem.

However  most mathematicians and physicists ignore any limitations imposed by
constructivist viewpoints.  Their activities appear to be based implicitly on the ideal
or Platonic viewpoint of mathematical existence, i.e. that mathematical entities and
statements have an ideal existence and truth status independent of any physical
limitations$^{(}$\cite{Penrose}$^{)}$ or an observers knowledge of
them$^{(}$\cite{Hersh,Kline}$^{)}$. The "luscious jungle
flora"$^{(}$\cite{Frankel}$^{)}$ aspect of this view of mathematical existence compared
to the more ascetic landscape$^{(}$\cite{Frankel}$^{)}$ of more constructivist views is
hard to resist.

However, this viewpoint has the problem that  one must face the
existence of two types of objects.  There are the ideal
mathematical objects that exist outside space-time and the
physical objects that exist inside of and influence the properties
of space-time.  The existence of two types of objects that appear
to be unrelated yet are also closely related is quite
unsatisfactory.

Another approach to mathematics is that of the formalist
school$^{(}$\cite{Hersh,Kline}$^{)}$, Here mathematics is  considered to be in essence
like a game in which symbol strings (statements or formulas) are manipulated according
to well defined rules.  The goal is the rigorous proof of theorems. Mathematical
entities have no independent reality status or meaning.

In one sense the formalist school is related to physics in that provability and
computability are closely related.  In work on computability and computational
complexity$^{(}$\cite{Papadimitriou}$^{)}$ it is clearly realized that computability is
related closely to what can be carried out (in an ideal sense) on a physical computer.
Yet, as has been noted$^{(}$\cite{DeutschMLQ}$^{)}$, the exact nature of the
relationship between computations carried out by real physical computers and abstract
ideal computers, such as Turing machines, is not clear.

The influence of physics on mathematics is perhaps  most apparent in recent work on
quantum information theory and quantum computing.  Here it has been
shown$^{(}$\cite{Shor,Grover}$^{)}$ that there exist problems that can in principle be
solved more efficiently on a quantum computer than by any known classical computational
algorithm. Also the increased efficiency of simulation of physical quantum systems on
quantum computers$^{(}$\cite{Feynman,Zalka,Wiesner,Abrams,Lloyd}$^{)}$ compared to
simulation on classical systems is relevant to these considerations.

The problems on the relationship between physics and mathematics have been considered
by others. In his insistence that "Information is Physical"
Landauer$^{(}$\cite{Landauer}$^{)}$ also recognizes the importance of this
relationship.  His reference to the fact that, according to Bridgman, mathematics
should be confined to what are in essence programmable sequences of operations, or that
mathematics is empirical$^{(}$\cite{Bridgman}$^{)}$, supports this viewpoint. Similar
views on the need for an operational characterization of physical and set theoretic
entities has been expressed$^{(}$\cite{Svozil}$^{)}$.

Other attempts to show the importance of physics on the foundations of mathematics
include work on randomness$^{(}$\cite{BenRAN}$^{)}$ and on quantum set
theory$^{(}$\cite{Finkelstein,Takeuti}$^{)}$ (see also$^{(}$\cite{Nishimura}$^{)}$).
Recent work on the relationship between the Riemann hypothesis and aspects of quantum
mechanics$^{(}$\cite{Odlyzko,Crandall}$^{)}$ and relativity$^{(}$\cite{Okubo}$^{)}$,
and efforts to connect quantum mechanics and quantum computing with logic, languages,
and different aspects of physics should be
noted$^{(}$\cite{Ozhigov,Buhrman,Schmidhuber,Blaha}$^{)}$ along with efforts to connect
mathematical logic with physics$^{(}$\cite{Tegmark,Spector,Foschini}$^{)}$.

The existence of  axiomatizations of physical theories in the literature also suggests
the significance of foundational aspects in the relation between physics and
mathematics. In particular axiomatizations of quantum mechanics and quantum field
theory have been much studied. These include algebraic
approaches$^{(}$\cite{Mackey,Haag}$^{)}$, quantum logic
approaches$^{(}$\cite{BirkhfVnN,Jauch}$^{)}$ and others$^{(}$\cite{Hardy}$^{)}$. These
axiomatizations are often quite mathematical and rigorous, but they refer to specific
physical theories and not to a general theory of physics and mathematics.

In spite of this progress both the lack of and a need for a coherent theory of
mathematics and physics together remain.  One view of this is expressed by the title of
a paper by Wigner$^{(}$\cite{Wigner}$^{)}$ "On the unreasonable effectiveness of
mathematics in the natural sciences". One does not know why mathematics is so effective
and an explanation is needed. Even though the paper was published in 1960, it is still
relevant today.  A related question, "Why is the physical world so
comprehensible?"$^{(}$\cite{Davies}$^{)}$ also needs to be answered.

These questions are still relevant today. It is hoped that the
following material will help to answer these questions by
providing part of a framework that can be used to construct a
coherent theory of mathematics and physics together.
\\

\section{A Coherent Theory of Physics and Mathematics}
\label{CTMP}  The basic idea is that a coherent theory of
mathematics and physics must be a description of both the
mathematical and physical components together of the universe.
They should not be treated as separate universes as has been done
so far. Whether such a theory would include all or just some
mathematics is not known, and one suspects that only some
mathematics would be included. However it is reasonable to require
that mathematical systems that are used or are potentially usable
by physics should be included in the domain of a coherent theory.

At present it is not known how to construct a coherent theory of mathematics and
physics. One purpose of this paper is to suggest that a way to accomplish this is to
work towards a combination of mathematical logical aspects with physics at a
foundational level. This approach, which is implied by other
work$^{(}$\cite{Tegmark}$^{)}$, seems worthwhile since mathematical logic is the study
of basic aspects of mathematical systems in general. This includes a study of
properties such as consistency, completeness, truth, and validity. By combining
mathematical logic with physics one may hope to  make these properties an integral part
of physics at a basic level. Furthermore such a combination may even influence what is
true in physics at a basic level.

A basic and possibly defining requirement of a coherent theory is
that it must be valid and sufficiently strong, and it must be able
to maximally describe both its own validity and sufficient
strength. Much of this paper is devoted to discussing this
requirement and showing why it may be quite important.

\subsection{\textbf{Validity and Sufficient Strength}}
It is first necessary to discuss properties that any satisfactory first order
axiomatizable theory should have$^{(}$\cite{Shoenfield,Smullyan}$^{)}$. Some of this
subsection is a review of well known material. A good theory should be such that all
properties predicted by the theory should be true in its domains of applicability. Also
the theory should be sufficiently strong in terms of its predictive power to be
regarded as a satisfactory or useful theory.

Since theories differ by having different sets of axioms, the interest is in the
properties of a sets of axioms for theories, including a coherent theory of physics and
mathematics. One property that a theory must have is that it is valid. A theory is
defined to be {\em valid} for a structure$^{(}$\cite{Shoenfield}$^{)}$ if all theorems
of the theory are true in the structure.  For a coherent theory the structure is that
part of the physical and mathematical universe to which the theory applies (this will
be discussed more later on) and truth in the structure has the usual intuitive informal
meaning. Following the usual practice in mathematical logic, no definition of truth is
provided as it is an intuitive concept\footnote{As is well known, one can define the
truth of formulas in a language of a theory  by induction on the logical depth of the
formulas. However the definition depends on the intuitive notion that is used to define
the truth of the atomic formulas at the base of the induction.} Even so various
properties of the set of true formulas of theories such as arithmetic can be
derived$^{(}$\cite{Smullyan}$^{)}$.

The requirement of validity is essential as it connects the formal notion of proof and
theoremhood of a theory with the notion of truth in a structure.  But it is not
sufficient.  To see this note that the definition of validity says that for all
formulas $F$ in the language of the theory, if $F$ is a theorem, then it is true. The
problem is that this definition admits, as valid, theories that are too weak.  Consider
for example an empty theory with no (nonlogical or logical) axioms. Since it has no
theorems it is valid.  This follows from the truth table of if-then statements in that
such statements are true if the "if" part is false. Another example of a theory that is
too weak is first order predicate calculus as a theory with just logical axioms and no
nonlogical axioms.  This theory is valid as all the theorems are logical consequences
of tautologies$^{(}$\cite{Shoenfield}$^{)}$ and are true in any universe.  The logical
axioms would be present in any axiomatization of a coherent theory.

As is well known in mathematical logic, weak theories can be
strengthened by adding more nonlogical axioms. Extension of a
theory by adding more axioms gives a theory that is stronger in
that it has more theorems and predictive power than the unextended
theory. However, if the added axioms are just axioms that define
new symbols for existing terms in the language of the old theory,
then the new theory is not stronger than the old.

These considerations show that one must require that a coherent
theory is sufficiently strong.  This means that the set of axioms
should be sufficiently large and appropriate so that the resulting
theory will be recognized as a suitable coherent theory of
mathematics and physics.

It must be emphasized that this is not a definition of sufficient
strength.  It is not known at present how to give a definition.
However it can be shown that the idea has merit and is in fact
already in use for incomplete theories such as arithmetic and set
theory.

The first question to consider is whether a coherent theory will be complete or
incomplete. A theory is complete if for each closed formula (a formula with no free
variables) in the language of the theory, either the formula or its negation (but not
both if the theory is consistent) is a theorem of the
theory$^{(}$\cite{Shoenfield}$^{)}$. A complete theory is of maximum strength. There is
no way to increase its strength by adding more axioms. Examples of complete theories
include those for the real numbers as a real closed field and atomless Boolean
algebras$^{(}$\cite{Shoenfield,Keisler}$^{)}$.

As was first shown by G\"{o}del$^{(}$\cite{Godel,Smullyan}$^{)}$ there are also
incomplete theories.  Good examples are arithmetic and any other theory, such as
Zermelo Frankel Set theory, sufficiently strong to include arithmetic. G\"{o}del first
showed incompleteness by exhibiting a formula that asserted its own unprovability.  If
arithmetic is valid, then neither this formula nor its negation can be a theorem of
arithmetic.

This argument was later extended by Chaitin$^{(}$\cite{Chaitin}$^{)}$ to show that
arithmetic, and any axiom system that includes the natural numbers, is quite
incomplete. In particular it follows from his work that almost all formulas that
express the randomness or algorithmic complexity of numbers are not theorems of
arithmetic. This follows from the observation that almost all length $n$ bit strings $
\underline{s}$ are random in that the algorithmic complexity of $ \underline{s}$ is
equal to $n + c$ which is the length $m$ in bits of the shortest program that generates
$ \underline{s}$ as output. Here $c$ is a constant. Also the number of random
$\underline{s}$ increases exponentially with $n$. In addition Chaitin's incompleteness
theorem$^{(}$\cite{Chaitin}$^{)}$ states that no formal axiom system whose axioms have
algorithmic complexity $p$ can be used to prove formulas stating that a specific
$\underline{s}$ has complexity $m$ with $m\geq p$. Proofs of complexity or randomness
are limited to statements about complexity values $<p$.

Since arithmetic is so incomplete it follows that there is much room for extending the
strength of arithmetic in may ways.  One such path is the nonterminating extension
provided by G\"{o}del's second incompleteness theorem$^{(}$\cite{Godel,Smullyan}$^{)}$.
This theorem says that in any theory  $T$ strong enough to express by a formula in the
language of $T$, the consistency of $T$, then that formula is not a theorem of $T$.
Also if one strengthens $T$ by adding new axioms so that the consistency of $T$ can be
proved in the stronger theory, then the same incompleteness result hold for the
stronger theory.

Besides this example there are many other ways or paths to follow
in strengthening arithmetic by adding new axioms. In view of this
it is surprising that the relatively simple axiomatization of
arithmetic, that is in use, is sufficiently strong for most uses
made of the theory. The reason, which is in many ways remarkable,
is that arithmetic, and mathematics in general and physics is
really concerned with small numbers with complexities of at most a
few hundred bits and with arithmetic operations on these numbers.
Really large numbers, such as those of the order of $2^{n}$ where
$n=10^{20}$ with correspondingly large complexities, seem to play
no role in arithmetic or physics.

Zermelo Frankel set theory with the axiom of choice (ZFC) is also incomplete.  It is
also powerful enough to include all the mathematics used so far by physics. However,
unlike the case with arithmetic, changes in the axioms of the theory have been
considered that depend on how the theory is to be used. One example is the extension of
the theory to include proper classes$^{(}$\cite{Frankel}$^{)}$ to include objects that
are not sets, such as the class of all sets that are not members of themselves and the
class of all ordinals.

Other changes are based on the existence of several easily formulated mathematical
statements whose truth value is unknown. The main example is the continuum hypothesis
(CH) which has been shown to be independent of the other axioms of
ZFC$^{(}$\cite{Cohen}$^{)}$. Thus one can study the properties of ZFC plus CH and of
ZFC plus the negation of CH. Other ways of extending or changing ZFC include the
addition of large cardinal axioms or replacing the axiom of choice by the axiom of
projective determinacy$^{(}$\cite{Woodin}$^{)}$.  A recent review of some of these
aspects is provided in$^{(}$\cite{Stillwell}$^{)}$.

This and other work on arithmetic and set theory shows that these
incomplete theories, axiomatized as Peano arithmetic and as ZFC
set theory, are sufficiently strong for almost all uses in
mathematics and physics even though they are quite incomplete.
These results should also apply to a coherent theory of physics
and mathematics. Any axiomatization of a coherent theory is
expected to be incomplete as arithmetic is included as part of
the mathematical component. A coherent theory  should also be
valid and be sufficiently strong for almost all uses.

These arguments all assume that it is possible to axiomatize a
coherent theory of physics and mathematics.   Whether this is
possible or not, or if there are many different coherent theories
each with their own axiom sets, will have to await further work.

Other aspects of the requirement of validity and sufficient
strength for a coherent theory  need discussion. It was noted
that validity means that all formulas in the language of a
coherent theory that are theorems must be true in the domain of
applicability of the theory, which includes physical and
mathematical systems. Applied to physical systems this means that
all properties of physical systems that are predicted by theorems
of the theory and are capable of experimental verification or
refutation, are verifiable by experiment. Sufficient strength
means that there must be sufficiently many properties of physical
systems that are predicted by theorems of the theory and can be
experimentally tested.

One should emphasize the role that physical procedures play in
this requirement. The requirement of validity means that all
theoretically predictable properties of systems that are
experimentally testable, are true.  Experimental testability of
any prediction requires the existence of physical procedures for
preparing a system in a specified state and for carrying out the
required measurement.

This requirement is problematic for quantum systems in that there
is no way so far to define physical implementability for
preparation of quantum systems in various states and for
measuring observables, represented by self adjoint operators in
an algebra of operators. It may well be that there are many
properties of systems that are predicted to be true for physical
systems,  but there do not exist any corresponding experimental
physical procedures for actually testing the prediction that are
implementable in an efficient manner.

For example, it is clear that for many states of complex quantum systems and for many
observables it is very unlikely that there exist efficiently implementable physical
procedures for preparing the states and measuring the observables. Examples of these
states include complex entangled states of multicomponent systems of the type studied
by Bennett et al$^{(}$\cite{Bennettetal}$^{)}$.  Examples of observables include
projection operators on these entangled states. A related example is based on the
observation that efficient physical implementability is not preserved under arbitrary
unitary transformations of self adjoint operators$^{(}$\cite{BenEIPSRN}$^{)}$. Thus if
the observable $ \check{O}$ is efficiently implementable it does not follow that
$U\check{O}U^{\dagger}$ is physically implementable for arbitrary unitary $U$. It has
also been noted that there is a problem in determining exactly which logical procedures
or algorithms are physically implementable$^{(}$\cite{DeutschMLQ}$^{)}$. This problem
is especially relevant for quantum computer algorithms as it is not at all clear which
are efficiently physically implementable and which are not.

The same requirement of physical implementability applies to computers as physical
realizations of  any one of the equivalent representations of abstract computability. A
much studied and quite useful abstract concept is that of Turing machines$^{(}$\cite
{Turing}$^{)}$.  These machines, described by a multistate head that moves along a tape
of cells and interacts locally with the cells, are both quite simple and very powerful.
In particular, by the Church Turing thesis, any computable function is computable by a
Turing machine.  So there are Turing machine equivalents for any existing computer.
Also Turing machines provide a convenient venue for proving the unsolvability of the
halting problems and the existence of universal computers, both quantum and
macroscopic$^{(}$\cite{Turing,Deutsch,Bernstein}$^{)}$.

As is well known, limitations of the computation process make many calculations  of
properties of systems predicted by theory very hard or impossible. In this case various
model assumptions and simplifications are used to make the calculations more tractable.
The use of quantum computers, if and when they are developed, to make some of these
calculations may help in that some problems$^{(}$\cite{Shor,Grover}$^{)}$ are much more
tractable on quantum computers than on classical computers.  The observation that
quantum computers are much more efficient at simulating the behavior of quantum systems
than are classical computers$^{(}$\cite{Abrams,Lloyd}$^{)}$ may also help.

\subsection{\textbf{Maximal Description of Validity and Sufficient
Strength}} \label{MDVSS}

It was noted earlier that it is not known if it is possible to
axiomatize a coherent theory or if there are many axiomatizable
coherent theories.  The other part of the basic requirement of a
coherent theory, that it describe to the maximum extent possible
its own validity and sufficient strength, may be relevant to these
considerations.

The meaning of this requirement is that there are one or more
formulas in the language of a coherent theory that can be
interpreted to express to some extent that the theory is itself
valid and sufficiently strong. In addition these formulas must be
true. This condition is expressed by the requirement that a
coherent theory is valid and sufficiently strong. It is also
desirable, but unlikely, that these formulas be theorems of the
coherent theory they are describing; one way of approaching this
aspect by theory extension will be discussed soon.

Now it may be the case that there are many candidate coherent
theories, each described by a different set of nonlogical axioms,
that differ in their ability to express their own validity and
sufficient strength. In some the true formulas express very little
or none of the validity and strength of the theory.  In other
coherent theories there may be true formulas that express more of
the validity and strength of the theories.

The condition of description of their own validity and strength
{\em to the maximum extent possible} serves an essential role in
that it limits or restricts the choice of candidate theories. Only
those candidate theories that maximally describe their own
validity and strength are acceptable as coherent theories.
However, use of this maximality condition to limit or restrict the
candidate theories, requires that one have a measure of the extent
of a theory's description of its own validity and sufficient
strength. Whether such a measure exists and, if so whether it has
a maximum or not, are questions for the future.

It was noted that formulas of a coherent theory that describe to
some extent the validity and sufficient strength of the theory
must be true. In addition one would like them to be theorems of
the theory.  In view of G\"{o}del's second incompleteness theorem
this is unlikely.  However the theorem suggests a possible way out
that throws new light on the condition that a theory {\em
maximally} describe its own validity and sufficient strength.

To begin it is expected that in any coherent theory in which it is possible to express
the maximal validity and sufficient strength of the theory, then the formula or
formulas expressing it are not theorems of the theory. The reason is that expression of
validity of a theory is similar to the expression of consistency when G\"{o}del's
completeness theorem is taken into account. (A theory is consistent if and only if it
has a model$^{(}$\cite{Shoenfield}$^{)}$.) If such a coherent theory $T_{0}$ exists
then it may be possible to extend and strengthen the theory by adding axioms so that
the formula expressing the  validity and sufficient strength of $T_{0}$ is a theorem of
the extended theory $T_{1}$.  But then the formula or formulas expressing the  validity
and sufficient strength of $T_{1}$ are not theorems of $T_{1}$.

This suggests the consideration of an iterated extension of
theories where the formula or formulas $\{F_{n}\}$ expressing the
validity of $T_{n}$ are theorems in $T_{n+1}$ but the formulas
$\{F_{n+1}\}$ are not theorems of $T_{n+1}$.  Since there is no
upper bound to $n$, one is faced with a nonterminating extension
of theories. It may also be the case that there are many possible
iterated extensions, with the theorems of $T_{n+1}$ describing to
some degree the validity and strength of $T_{n}$. In this case one
limits the paths or extensions by requiring that at each stage in
the extension $T_{n+1}$ must be such that the formulas that {\em
maximally} describe the validity of $T_{n}$ are theorems of
$T_{n+1}$.  In this way the condition of {\em maximal} description
limits or restricts the extension paths.

In an even more speculative vein, it may even be possible to at
least partly axiomatize the extension process itself. In this way
one might hope to obtain one or a few coherent theories that
describe to the maximum extent possible the process of extension
of parts of themselves. Each of these "limit coherent theories"
would describe a nonterminating extension of parts of themselves.
It might be that the qualification "to the maximum extent
possible" is very useful in restricting an iterated selection
process that leads to or selects a "limit coherent theory". For
example it may be that the theory at each stage of the iterated
extension is algorithmically more complex than the theories at
earlier stages. The theory at each stage may also maximally
describe its own validity and sufficient strength by maximally
describing and proving the corresponding formulas for the
theories in all preceding stages\footnote{One wonders if the
maximal axiomatization of this extension process, as a
restriction on the possible paths of extension, has any overlap
with the least action principle used in physics to restrict the
paths taken by systems as described by Feynman path integrals.}.

It is easy to dismiss all of the above as idle speculation with no
basis in fact.  This may indeed be the case.  However it is worth
noting that a coherent theory of physics and mathematics, by
virtue of including both physical systems in space-time and
mathematical systems in its domain, is quite different and may be
potentially more powerful than is a purely mathematical theory
whose domain is limited to mathematical systems. It is hoped that
the reader will appreciate this possibility after reading the
following sections, especially Sections \ref{EPMU} and \ref{LP}.

\subsection{\textbf{Universal Applicability}}
Another distinguishing property of a coherent theory is that the
theory should be universally applicable. This means that  the
domain of applicability should include a "sufficiently large"
component of the physical and mathematical universes.   These two
requirements, inclusion of both physical and mathematical systems
and universal applicability, might be expected to yield some new
results not obtainable from purely mathematical theories or purely
physical theories that ignore foundational aspects of mathematics.
It will be seen later that this may be the case.

At this point the precise definition of "sufficiently large" must be left to future
work.  However, it is worth repeating that, as is well known, there is no single theory
that is universally applicable to all of mathematics. One such attempt, Zermelo Frankel
set theory$^{(}$\cite{Frankel}$^{)}$, foundered on the Russell antimony such as the
"set of all sets that do not contain themselves", and other similar "sets". Extensions
of set theory to include proper classes$^{(}$\cite{Frankel}$^{)}$ may solve these
problems but they leave untouched other aspects.

It is also not clear at this point how to exactly define universal applicability for
the physical component of the theory, or whether one should even separate the theory
into mathematical and physical components.  In the case of quantum mechanics, one
view$^{(}$\cite{Tegmark1}$^{)}$ is that defining universal applicability to include all
physical systems means that one must accept the Everett
interpretation$^{(}$\cite{Everett,Wheeler1}$^{)}$ of quantum mechanics. This
interpretation assumes that the whole universe is described as a closed system by a
quantum state evolving according to quantum dynamical laws.

One way to avoid this may be to assume that universal
applicability means that the theory is applicable only to systems
that are subsystems or part of other larger systems that are in
turn subsystems of other still larger systems.\footnote{A more
precise statement of this might be: (1) If the theory is
applicable to subsystem $A$, then there exist many subsystems $B$
that contain $A$ and to which the theory is applicable. (2) There
are many subsystems $A$ to which the theory is applicable.
Furthermore the definition of "many subsystems" must be
sufficiently broad to include all subsystems accessible to state
preparation and experiment.} This includes both open and closed
subsystems including those that may be isolated for a period of
time.

At present it is not clear if an exact definition is needed.
However a definition should be such to include subsystems
described by a finite number of degrees of freedom and many
systems, such as quantum fields, that are described by an infinite
number of degrees of freedom.  One should not be able to point to
or describe a physical system occupying a finite region of space
and time that is not included in the theory.

An essential aspect of this is that intelligent systems should be
included in the domain of a coherent theory.  If quantum mechanics
is universally applicable it then follows that intelligent systems
are quantum mechanical systems.  The observation that the only
known examples (including the readers of this paper) of
intelligent systems are macroscopic, with about $10^{25}$ degrees
of freedom, does not contradict the quantum mechanical nature of
these systems. This may well be a reflection of the possibility
that a necessary requirement for a quantum system to be
intelligent is that it is macroscopic. However, whether this is or
is not the case, is not known at present.

That intelligent observers are  both conscious self aware systems and quantum systems
has been the basis for much discussion on consciousness in quantum
mechanics$^{(}$\cite{Penrose,Stapp,Squires,Page}$^{)}$. Included are discussions on
interactions between two quantum observers$^{(}$\cite{Wigner1,Albert}$^{)}$. These
avenues will not be pursued here as they do not seem to be the best way to progress
towards developing a coherent theory of mathematics and physics.

It follows that the dynamics of the quantum systems carrying out
the validation of quantum mechanics must be described by quantum
dynamical laws.  Thus quantum mechanics must be able to describe
the dynamics of its own validation process. However validation of
a theory involves more than just describing the dynamics of the
systems carrying out the validation.  Validation includes the
association of meaning to the results of theoretical derivations
and computations carried out by quantum systems (as computers).
Meaning must also be associated to the results of carrying out
experiments by quantum systems (as robots or intelligent systems).

This association of meaning to the results of quantum processes is
essential.  It is basic to determining which processes are either
computational or experimental procedures.  These processes are a
very small fraction of the totality of all processes that can be
carried out, most of which have no meaning at all.  They are
neither computations or experiments.

This association of meaning includes such essentials as the (nontrivial) assignment of
numbers to the results of both computational and experimental process. A computation
process or an experiment that halts  produces a complex physical system in a particular
physical state.  What numbers, if any that are associated to the states depend on the
meaning or interpretation of the process$^{(}$\cite{BenRNQM,BenRNQMALG}$^{)}$.  That
is, the process must be a computational or experimental procedure.  If it is, then one
must know the property to which it refers. This is needed to know the association
between theoretical computations and experimental procedures.

For example in quantum mechanics for some observable $\check{O}$
and state $\Psi$,  one must be able to determine which procedure
is a computation of the expectation value $\langle
\Psi|\check{O}|\Psi\rangle$. One must also know which experimental
procedure corresponds to a measurement of this expectation value.
This assumes that requisite experimental and theoretical
calculation procedures exist for $\check{O}$ and  $\Psi$. As is
well known the experiment must in general be repeated may times to
generate the expectation value as a limit as $n\rightarrow \infty$
of the average of the first $n$ repetitions of the experiment.
Association of meaning to these procedures also includes all the
components involved in determining that appropriate limits exist
for both the computation procedures  and experimental procedures.

A coherent theory of mathematics and quantum mechanics must be
able to express as much of this meaning as is possible.  Not only
must it be able to express the dynamics of its own validation,
but it must be able to express the meaning associations described
above to the maximum extent possible. This is part of being able
to maximally describe its own validation.

A similar argument applies to sufficient strength.  That is,
there must be a sense in which a coherent theory is sufficiently
strong to include all properties that are predictable and capable
of experimental test. Of course the problem here lies in the exact
meaning of "all properties that $\cdots$". One may hope and
expect that a coherent theory would be able to express to the
maximum extent possible the meaning of "all properties that
$\cdots$". And it should also be able to express the condition
that it is also sufficiently strong.

These arguments are part of the basis for the requirement that the
coherent theory maximally describe its own validation and
sufficient strength.  However, being able to generate such a
description does not guarantee that the coherent theory {\it is}
valid and sufficiently strong. A theory may be interpreted to
express that it has some property, but it does not follow that it
actually has that property. This possibility is taken care of by
the other component of the basic requirement of a coherent theory,
that it is valid and sufficiently strong.

\subsection{\textbf{A Coherent Theory and the Strong Anthropic Principle}}

The conditions that a coherent theory include both physics and
mathematics and that it satisfy the requirement of maximal
description of its validity and sufficient stremgth and be
maximally valid and sufficiently strong, suggest that there may
be a very close relation between the theory and the basic
properties of the physical universe.  It may be the case that at a
very basic level the basic properties of the physical universe are
entwined with and may even be determined by a coherent theory that
satisfies the requirements.

Examples of such basic properties  that may emerge from or be determined by the
coherent theory include such aspects as the reason for three space and one time
dimension (See Tegmark$^{(}$\cite{Tegmark}$^{)}$ for another viewpoint), the strengths
and reason for existence of the four basic forces, why quantum mechanics is the valid
physical theory, etc.. Even if few or none of these properties are determined, one may
hope that the theory will shed new light on already explained basic properties.

These possibilities suggest that a coherent theory with the requirement is related to
the strong anthropic principle$^{(}$\cite{BarTip,Hogan,Greenstein}$^{)}$. This
principle can be stated in different ways. One statement is that "The basic properties
of the universe must be such that [intelligent]life can
develop"$^{(}$\cite{BarTip}$^{)}$. Wheeler's interpretation as quoted by Barrow and
Tipler$^{(}$\cite{BarTip}$^{)}$ is that "Observers are necessary to bring the universe
into being".  A stronger statement is the final anthropic
principle$^{(}$\cite{BarTip}$^{)}$ "Intelligent information processing must occur and
never die out".

The relation between this principle and a coherent theory can be
seen by recasting the statement of the maximal validity and
sufficient strength requirement into an existence statement or
condition: {\em There exists a coherent theory of physics and
mathematics that is valid and sufficiently strong and maximally
describes its own validity and sufficient strength}. In this
case the basic properties of the physical universe emerge from or
are a consequence of the existence statement. That is, the basic
properties of the physical universe must be such that the
existence statement is true.

Another way to state this is that the basic properties of the
physical universe must be such that a coherent theory is
creatable. Since intelligent beings are necessary to create such a
theory, it follows that the basic properties of the physical
universe must be such as to make it possible for intelligent
beings to exist. Since the intelligent beings, as physical
systems, are part of the physical universe, the theory must, in
some sense, also refer to its own creatability.

None of this implies that intelligent beings must exist, only that
it must be possible for them to exist. Of course existence of
intelligent beings is a necessary condition for the actual
creation of such a coherent theory.

\subsection{\textbf{The Possible Uniqueness of a Coherent Theory}}
\label{PUCT} The requirement that a coherent theory of
mathematics and physics is valid and sufficiently strong and
maximally describe its own validity and sufficient strength
would seem to be quite restrictive. Indeed one may speculate that
the condition is so restrictive that there is just one such
theory.

One reason this might be the case is that if there were several
different coherent theories each satisfying the requirement, then
there might be several different physical universes, with the
basic physical properties of each universe determined by one of
the theories.  Yet we are aware of just one physical universe, the
one we inhabit, with the basic properties determined by both
physical theory and experiment. It follows that if the basic
properties of the physical universe are determined by a coherent
theory satisfying the requirement, then the existence of just one
physical universe implies that there is just one coherent theory
satisfying the requirement.

Viewed from this uniqueness perspective, the basic statement that
there exists just one coherent theory of physics and mathematics
that is valid and sufficiently strong and maximally describes its
own validity and sufficient strength becomes a quite powerful
restrictive condition. The reason is that it can be used with the
arguments given above to obtain the result that there is just one
physical universe with basic properties determined by the unique
theory. And this should be our universe.

This possibility was discussed earlier as a restriction on the
process of iterated extensions of a theory where the theory at
each stage of the iteration was required to be such that  formulas
in the theory maximally described the validity and sufficient
strength of the theories in the prior stages.  Also the
possibility of  "limit coherent theories" was considered. Here the
possibility that this restriction results in just one coherent
theory is being considered.

If this line of reasoning is indeed valid and there exists just
one coherent theory satisfying the basic requirement, then it
would be very satisfying as it answers the question, "Why does our
physical universe have the properties it does?". Answer: The
physical universe could not be otherwise as it is the only one
whose properties emerge from or are determined by and determine a
coherent theory. No other universe is possible because there is
just one coherent theory satisfying the maximality requirement and
each such theory is associated with just one physical universe.

At present this argument, although appealing, must be regarded as
speculation. Whether it is true or not must await development of a
coherent theory of physics and mathematics, if such is even
possible.
\\

\section{Emergence of the Physical and Mathematical Universes} \label{EPMU}
At present the main approach to physics seems to be that one
assumes implicitly a physical universe whose basic properties
exist independent of and a priori to a theoretical description,
supported by experiment, of the universe. This is implied by
reference to experiments as "discovering properties of nature". An
a priori, independent existence of the physical universe is also
implied in the expression used above "theoretical description,
supported by experiment, of the universe".

The approach to mathematics is much more variable as there are many different
interpretations of the meaning of existence in
mathematics$^{(}$\cite{Hersh,Kline,MarMyc}$^{)}$.  However, the Platonic viewpoint that
is widely accepted, at least implicitly, is that mathematical objects exist a priori to
and independent of a theoretical description of them with their properties to be
discovered by mathematical research.

Here the position is taken that one should regard the basic
properties of the physical and mathematical universes as very much
entwined with a coherent theory of mathematics and physics.
Neither the mathematical universe, physical universe, nor the
coherent theory should be considered to be a priori and
independent of the other two components.  The basic properties of
all three components should be considered to be emergent together,
and mutually determined and entwined.

This means that, for the relation between the physical universe
and the coherent theory, the basic physical aspects of the
physical universe should be considered to emerge from and be
determined by the basic properties of a coherent theory of physics
and mathematics. Also the basic properties of a coherent theory
should, in turn, emerge from and be determined by the basic
properties of the physical universe.

It must be strongly emphasized that the emergence noted above does
{\em not} mean that there is any arbitrariness to the basic
physical properties and that an observer can choose them as he
pleases. Rather the viewpoint taken here suggests that a coherent
theory that is  valid and sufficiently strong and maximally
describes its own validity and sufficient strength, is also
maximally objective. The reason is that a maximally self
referential theory refers to as much of its own consistency,
validity, and strength  as is possible, and the role of an
observer or intelligent being is thereby minimized in determining
the basic properties of the theory.  In this case the basic
properties of the universe as described by a coherent theory must
appear to any observer to be objective and real and maximally
independent of the existence and activities of an observer. That
is what one means by objectivity.

\subsection{\textbf{Dependence of Reality Status on Theory}}

In one sense the idea of the emergence of basic physical
properties of the universe is already in use. This is based on the
observation that the more fundamental properties of the physical
universe require many layers of theory supported by experiment to
give them meaning. Their reality status is more indirect as it
depends on many layers of theory supported by experiment.

For example the existence and properties of atoms is indirect in
that it is based on all the experimental support for the many
theoretical predictions based on the assumed existence and
properties of atoms.  One does not directly observe individual
atoms.  Pictures of individual atoms taken with an electron
microscope depend on many layers of theory and experiment to
determine that a complex physical system is an electron microscope
and that the output patterns of light and dark  shown on film or a
screen  are not meaningless but have meaning as pictures of
individual atoms.

The physical reality and properties of more fundamental systems,
such as quarks and gluons, are even more indirect than for atoms
and depend on more intervening layers of theory and experiment.
The same holds for neutrinos as fundamental systems whose reality
status and properties are quite indirect. Experimental support for
the existence of these particles depends on the layers of theory,
which may include quantum electrodynamics, and all the supporting
experiments needed to describe the proper functioning of large
particle detectors and assigning meaning to the output of the
detectors.

A similar situation exists for large, far away objects such as
quasars. The reality status and physical properties of these
systems are based on the theories of relativity and interactions
of electromagnetic fields, etc.. These are needed to interpret the
observations made using telescopes and to describe the proper
functioning of telescopes and other equipment used.

On the other hand the reality status and some properties of other
physical systems require little or no theoretical or experimental
support.  For example, the existence and hardness of rocks or the
existence of the sun and the facts that it is hot, bright and
round, are directly observed properties. Little theory with
supporting experiment is needed to make these observations. Other
properties of these objects are more indirect. An example is the
description of the sun as a gravitating body generating energy by
thermonuclear fusion of hydrogen.

It should be emphasized that none of the above implies that
systems such as quarks, atoms, and quasars are any less real and
objective than  are rocks and the sun.  Rather the point is that
the reality of their existence and their properties are more
indirect in that they depend on more intervening layers of theory
and experiment than is the case for rocks and the sun. Also the
reality of all the properties of quarks, atoms and quasars, is
indirect in its dependence on layers of theory and supporting
experiment.  For nearby moderate sized objects some of the
properties are quite direct and some are more indirect.  For
example, as noted above, direct properties of the sun are that it
is hot bright and round.  Indirect properties include the source
of its energy.

It is necessary to emphasize the importance of the intervening
layers of theory and experiment needed to support the proper
functioning and interpretation of complex equipment. Since most
equipment involves the electromagnetic interactions between
systems or between fields and systems, the theory of these
interactions must be well understood to ensure that a given
physical system is a properly functioning  piece of equipment.
This is needed to to ensure that certain properties of the system
represent output and that the output has meaning.
\\

\section{Language is Physical}\label{LP}
Additional support for the close relation between a coherent
theory and physics and mathematics is based on the essential
nature of language.  This applies to formal languages such as
those studied in mathematical logic and informal languages such
as English that are used for communication or transmission of
information among intelligent beings or for thinking.

The essential point to make here is that language is physical. All
language expressions must be represented by physical systems in
some states.  This applies irrespective of whether the language is
written or spoken and of how the basic units of the language are
organized. Depending on the physical representation the basic
units can be a set of symbols or characters organized into strings
as in written language, or a set of syllables appropriate for a
spoken language.

The importance of this aspect is emphasized by the observation
that, if it were not possible to represent language by states of
physical systems, it would not be possible to communicate or
acquire knowledge, or even think. It is an essential part of the
existence of intelligent observers, as language is an essential
part of the communication of information.

\subsection{\textbf{Physical Models of Language}}\label{PML}
There are many possible representations for languages based on
symbols, words, and word strings. Examples include printed text,
modulated waves moving through some medium as is the case for
spoken language, or language transmitted optically by use of
photons. As a more specific example, consider the text of this
paper. Each letter, word, paragraph, etc. is represented by
physical systems in different physical states. This is the case
whether the paper appears as printed material on pages of paper or
as patterns of light and dark regions on a computer screen.  If
the text is read, as in a lecture, the language is represented by
time variations in phase and amplitude of sound waves. It also
applies if the paper is represented as a large tensor product
state of quantum systems where each letter of the language is
represented by a state of a component quantum system in the tensor
product.

Some details of a representation of language text by arrangements
of ink molecules located on a two dimensional lattice of potential
wells are described in the Appendix. This representation is just
one of many possible. Another one, based on spin projection states
of systems, will be outlined below.  More generally, one can use
any physical observable with a discrete spectrum  and eigenstates
that can be associated with the language units such as symbols or
syllables. Also it must be possible to actually physically prepare
systems in these eigenstates and to measure the properties of
these systems corresponding to different properties of the
language text.

A simple way to  construct a quantum mechanical product state of a
language is based on the well known use of tensor products of
qubit states as a binary representation of numbers. Extension of
this model to $k-ary$ representations of numbers is done by
considering each component to be a qukit or $k$ dimensional
system.  The states of each qukit in some chosen basis represent
the $k$ digits or numerals of the representation. This can easily
be taken over to a language representation by letting the basis
states of each qukit correspond to the $k$ symbols in the alphabet
of the language. (From now on the basic language units will be
referred to as symbols.)

To be specific let the language alphabet consist of the $13$
symbols $0,1,c,v,f, r,=,\vee, \exists,\neg,(,),\#$. $0$ and $1$
are constants; $c,v,f,r$ are constant, variable, function and
relation symbols; $=$ is the equality symbol; $\vee,\exists,\neg$
are logical symbols for conjunction, exists, and not; $(,)$ are
left and right parentheses used to distinguish components of
words; and $\#$ is a spacer symbol. This language is generic and
can be used to construct a language suitable for any theory for
which a denumerable number of constants in the language suffices.
Details of how to implement this will not be done as it is not
relevant here.

Consider a 1 dimensional lattice of points $j=\cdots
-1,0,1,\cdots$ with a spin system located on each point $j$ of the
lattice.  The spin $\sigma$ of each system is such that $2\sigma
+1\geq 13$. If $I$ is a one-one map from the symbol set to the
spin projection eigenstates, then spin projection basis states of
the form $|m,j\rangle$ where $-\sigma\leq m \leq \sigma$ and $m$
is in the range set of $I$ correspond under $I$ to symbol states
at site $j$. That is, the symbol state $|s,j\rangle$ corresponds
to $|I(s),j\rangle$ where $s$ is any of the 13 symbols.

Different types of symbol string states will be referred to in the
following. Expression states correspond to multisymbol states of
the form $|\underline{s}, [a,b]\rangle =\otimes_{j=a}^{b}
|\underline{s}_{j},j\rangle$  Here $[a,b]$ is an interval of
points on the lattice and $ \underline{s}$ is a function from the
set of lattice points to the set of $13$ symbols with $
\underline{s}_{j}$ the value of $ \underline{s}$ at $j$.  Word
states are expression states with no spacers included.  They have
the form $|w,[a,b]\rangle = |\underline{s},[a,b]\rangle$ where
$\underline{s}_{j}\neq \#$ for $a\leq j\leq b$.  Formula states
are word states with the symbols combined according to specific
spelling rules. In English these would correspond to the words in
a dictionary. In formal languages studied in mathematical logic
they are sometimes referred to as well formed formulas.

Each of these types of expression states, $|\underline{s},
[a,b]\rangle$, corresponds, under $I$, to a product state of spin
projections on the interval $[a,b]$.  Since the lattice is
infinite the system has infinitely many degrees of freedom and is
best treated by quantum field theoretic methods. Here this will be
avoided by restricting the set of basis states to have the form
$|\underline{s}, [-\infty ,\infty]\rangle \equiv
|\underline{s}\rangle$ where at most a finite number of values of
$ \underline{s}$ are different from $\#$. In the spin projection
model this corresponds to an arbitrary but finite number of spin
systems on the lattice having spin projections different from that
corresponding to $I(\#)$. The Hilbert space spanned by this basis
is separable as there are denumerably many such basis states.

Because $\#$ denotes a spacer, one is limited to expression states
corresponding to an arbitrary but finite number of word states
where each word state is of finite length. If
$|\underline{\#},[a,b]\rangle$ denotes a spacer string state in
the interval $[a,b]$, then all the basis states can be written as
alternating word and spacer string states:
\begin{eqnarray}|\underline{s},[a,b]\rangle
=|\underline{\#},[-\infty,a]\rangle\otimes |w_{1},[a +1,\alpha_{1}]\rangle\otimes
|\underline{\#}[\alpha_{1}+1,\alpha_{2}] \rangle\otimes \nonumber \\
 |w_{2},[\alpha_{2}+1,\alpha_{3}]\rangle\otimes \cdots
 \otimes |w_{m},[\alpha_{2m}+1,b]\rangle\otimes
|\underline{\#},[b+1,+\infty]\rangle. \label{wdstr}
\end{eqnarray} This state consists of $m$ word states separated by
spacer string states with the lattice intervals occupied by each
word or spacer string state shown by the subscripted $\alpha$s.

Pure states of the lattice systems can be represented in general
as linear sums of word string states, Eq. \ref{wdstr}, where the
sums range over the lengths and number of  word and spacer string
states in $|\underline{s},[a.b]\rangle$. Sums over the interval
variables $a,b$ are also included.

\subsection{\textbf{Efficient Implementability}}\label{EI}
A basic requirement on any  physical model such as the spin
systems or the ink molecule systems to be a representation of a
language under some interpretation $I$ is that transformations
corresponding to basic syntactic operations must be efficiently
implementable. Efficient implementability means that there must
exist an actual physical procedure for carrying out each operation
and the operation must be efficient.  That is, the space-time and
thermodynamic resources required to implement the operations on
states with $n$ symbols must be polynomial in and not exponential
in $n$.

Syntactic operations are state transformations that create new
expression states or change existing ones where the dynamics of
the operations depends only on the syntactic properties of the
states as expression states of a language and on the physical
properties of the systems representing the symbols, words, and
word strings. The dynamics is independent of any meaning the
physical states might or might not have under some interpretation
as text with meaning on a language or as formula states in a
(consistent) theory described by axioms.

Examples of syntactic operations include concatenation of two
expression states, addition of a negation symbol state to some
formula state, generation or enumeration of formula states or term
states according to the rules for generating terms or formulas in
a language, enumeration of the theorems of a theory, etc.. These
operations depend only on the symbols represented by the model
states and how they are combined into expression states and word
string states. If the dynamics of the physical system does not
result in efficient implementation of these operations, then then
it is not an admissible model of the language.

It is useful to discuss efficient implementability in terms of a simple dynamical
model. Let $U_{M}(t) =e^{-iH_{M}t}$ denote a unitary dynamics of a multistate system
$M$ moving along a lattice $[-\infty,\infty ]$ and interacting locally with spin
systems at each lattice point under the action of a Hamiltonian $H_{M}$. If the
complete initial system state at time $0$, $|\underline{\#}\rangle|i,0\rangle$,
describes the lattice system with the spin projections at all sites in the state
corresponding to the spacer symbol, and $M$ in internal state $|i\rangle$ and at
lattice site $0$, then $\psi (t)=U_{M}(t) |\underline{\#}\rangle|i,0\rangle$ is the
state at time $t$. In this case $\psi(t)$ can be written as a linear sum over all
internal states and positions of $M$ and over all word string states of the form of Eq.
\ref{wdstr}$^{(}$\cite{BenDTVQM, BenUMLCQM}$^{)}$.

The requirement that $H_{M}$ is physically implementable means
that there must exist actual physical procedures, as part of the
experimental setup, that can be carried out to ensure that $H_{M}$
is the correct Hamiltonian description of the dynamics of $M$
interacting with the lattice spin system. Also it must be possible
to physically prepare $M$ and the spin lattice in the initial
state corresponding to $|\underline{\#},i,0\rangle$.

In general $U_{M}(t)$ describes the dynamics of $M$ starting at
position $0$ and generating linear superpositions of word string
states as linear superpositions of spin projection string states
of increasing length on the lattice.  $H_{M}$ may be such that $M$
can move in either direction.  Also $M$ may return at any time to
change the states of systems in a lattice interval and might even
return periodically to any lattice site to make changes.

The requirement that $H_{M}$ efficiently implements some operation
excludes this periodic return possibility. $H_{M}$ must be such
that the probability of a change in the spin projection state of a
system at any lattice site decreases as time increases and that
the state of lattice systems in any finite lattice region is
asymptotically stable.

The requirement of efficiency has two components.  One is
asymptotic stability and the other is the rate at which stability
is approached.  As noted, stability is necessary because random or
fluctuating changes in the states of spin systems in any lattice
interval mean that the spin projection states of systems in an
interval do not correspond to a specific word or formula in a
language. In this case all parts of the text could be changed at
any time irrespective of when they were generated.

The requirement of efficiency can now be defined in terms of the
rate at which  the states of systems in any lattice region
approach asymptotic stability. The system is asymptotically stable
under the dynamics $U_{M}(t)$ if the limit
\begin{equation}\langle\underline{\#},i,0|
P_{ \underline{s},[a,b]}(\infty)|\underline{\#},i,0\rangle \equiv
\lim_{t\rightarrow\infty}\langle\underline{\#},i,0|P_{
\underline{s},[a,b]}(t) |\underline{\#},i,0\rangle
\label{asymstab}
\end{equation} exists for each interval $[a,b]$
and each state $|\underline{s},[a,b]\rangle$. Here the Heisenberg
representation is used where
\begin{equation} P_{ \underline{s},[a,b]}(t) =
U_{M}^{\dagger}(t)P_{ \underline{s},[a,b]}U_{M}(t). \end{equation}
and $P_{ \underline{s}[a,b]}$ is the projection operator for the
state $|\underline{s},[a,b]\rangle$.

The existence of the limit is expressed by the statement that for
each $m$ there is a time $\tau$ such that for all
$t,t^{\prime}>\tau$
\begin{displaymath} |\langle\underline{\#},i,0|
P_{ \underline{s},[a,b]}(t) |\underline{\#},i,0\rangle  -
\langle\underline{\#},i,0| P_{ \underline{s},[a,b]}(t^{\prime})
|\underline{\#},i,0\rangle|   <2^{-m}. \end{displaymath} In
addition this must hold for all intervals $[a,b]$ and all $
\underline{s}$ in $[a,b]$, i.e. for all states
$|\underline{s},[a,b]\rangle =
\otimes_{j=a}^{b}|\underline{s}_{j},j\rangle$.

Let $n=b-a+1$ be the length of $ \underline{s}$. It follows from
the existence of the limit that for each $n,\underline{s},m$ there
exists a smallest $\tau\equiv\tau(n,\underline{s},m)$ that
satisfies the above expression. The definitions and properties of
the dynamical system give the result that
\begin{displaymath}\begin{array}{ccl}
\tau(n,\underline{s},m)\leq\tau(n,\underline{s},m^{\prime})&
\mbox{ if } & m<m^{\prime} \\
\tau(n,\underline{s},m)\leq\tau(n^{\prime},\underline{s^{\prime}},m)&
\mbox{if }& \{ \begin{array}{l} \mbox{$ \underline{s}$ is an
initial part of $ \underline{s^{\prime}}$}
\\ \mbox{and $n^{\prime}$ is the
length of $ \underline{s^{\prime}}.$}\end{array}
\end{array}\end{displaymath} The dependence on $ \underline{s}$ can be
removed by defining $\tau(n,m) = \max_{ \underline{s}:L(
\underline{s})=n}\tau(n, \underline{s},m)$ where the maximum is
over all $ \underline{s}$ whose length $L( \underline{s})=n$. From
the above one has $\tau(n,m)\leq \tau(n^{\prime},m^{\prime})$ if
$n<n^{\prime}$ or $m<m^{\prime}$. Also the definition of the limit
holds if $\tau(n,m)$ replaces $\tau$.

The requirement of efficiency can be expressed in terms of the $n$
dependence of $\tau(n,m)$ for each $m$. The dynamics $U_{M}(t)$ is
efficient if for each $m$, the dependence of $\tau(n,m)$ on $n$ is
polynomial in $n$, or $\tau(n,m) = K_{m}n^{\ell_{m}}$. The
dynamics is inefficient if the dependence is exponential in $n$,
or $\tau(n,m) = C_{m}2^{n^{\mu_{m}}}$. Here
$K_{m},C_{m},\ell_{m},\mu_{m}$ are $m$ dependent positive
constants with $\mu_{m}$ usually $\sim 1$. (The possible presence
of $\log n$ factors is ignored here.) In most applications the
requirement of efficiency also means that $\ell_{m}$ is not too
large, and is of order unity.

The above discussion has ignored the dependence of $\tau$ on $a$,
i.e. $\tau = \tau(n,a,\underline{s},m)$. Such a dependence exists
because the values of $\tau$ depend on where $ \underline{s}$ is
located in the lattice.  However this dependence, which can be
expressed by a location dependence of the constants $K_{m}=
K_{m,a}$ and $C_{m}=C_{m,a}$ does not affect the rate of
convergence.

So far $H_{M}$ has been required to be such that it is physically
implementable and that $\langle\underline{\#},i,0| P_{
\underline{s},[a,b]}(t) |\underline{\#},i,0\rangle$ is
asymptotically stable for any state $|\underline{s},[a,b]\rangle$
and that stability be approached efficiently.  However no specific
operation has been mentioned.

If $H_{M}$ is to efficiently generate a specific word string
state, $|\underline{X}\rangle$ where $ \underline{X}(j)=\#$ for
all $j$ outside an interval $[a,b]$, then one requires that the
dispersion of the limit probability $\langle\underline{\#},i,0|
P_{ \underline{X^{\prime}},[a,b]}(\infty)
|\underline{\#},i,0\rangle$ about $\langle\underline{\#},i,0| P_{
\underline{X},[a,b]}(\infty) |\underline{\#},i,0\rangle$ is small
for $|\underline{X^{\prime}},[a,b]\rangle \neq
|\underline{X}[a,b]\rangle$ and that $\langle\underline{\#},i,0|
P_{ \underline{X},[a,b]}(\infty) |\underline{\#},i,0\rangle =
1-\epsilon$ with $\epsilon$ small.

In this type of model, $H_{M}$ is required to be such that the
probability distribution is concentrated around a particular word
string. Whether $H_{M}$ has this property or not depends on the
physical model and Hamiltonians that are efficiently implementable
for the model. The ink molecules on a lattice (Appendix) are
likely to be a model of this type.   This is indicated by the wide
use of printing and typing. Implementation of such a model by a
dynamics that is unitary, if such is possible, requires the
presence of many auxiliary systems, including a supply of ink
molecules.

For microscopic physical systems, such as the spin projection
systems, it is less clear if this type of model giving a small
dispersion around just one state is efficiently implementable. In
this case it may be appropriate to consider models closer in
spirit to those considered in quantum computation. Then one
requires that $H_{M}$ is such that a linear superposition of word
string states is generated where the probability is high that each
component in the superposition has a specific property.

One example would be a theorem enumeration process for an
axiomatizable theory where each word string state in the
superposition is a string of formula states that corresponds to a
proof in the language of the theory. Each formula is either an
axiom (logical or nonlogical) of the theory or a formula derived
from an already generated formula in the string by application of
one of the logical rules of deduction. Thus the state generated by
$U_{M}(t)$ becomes a linear superposition of proof states in which
each formula state in each string is a theorem of the theory.
Since the enumeration process does not halt, the length of each
formula string state in the superposition increases with
increasing $t$.  Of course the existence of such a model depends
on the existence of a Hamiltonian such that the dynamics with the
required properties is efficiently implementable.

The usefulness of the simple model of a system $M$ moving along a lattice of systems
derives from the observation that there are several interesting examples of dynamical
systems generating output that may represent strings of words in some language.  Any
quantum system generating text, such as the writing of a research paper, is an example.
Another would be a complex quantum system, such as a quantum
robot$^{(}$\cite{BenQR}$^{)}$ or intelligent system, moving about in a complex
environment of quantum systems and generating output in the form described above.
Relevant questions for this example include "Does the output have meaning to us as
external observers?; If so, what is the meaning?". "Does the output also have meaning
to the system $M$ generating the output? If so, do the two meaning interpretations
coincide?"

A simple example of such a system $M$ moving along a one dimensional lattice initially
in the constant spacer state $|\underline{\#}\rangle$ and generating states that are
linear superpositions of word string states in a very simple language has been
described$^{(}$\cite{BenDTVQM,BenUMLCQM}$^{)}$. Here some word states were taken to
have meaning in that they denoted the appearance or nonappearance of other specified
word states in the word string states.

It was seen that the requirements that the dynamics of $M$ be
valid (any word that has meaning must be true) and complete (each
word with meaning must appear in some word string state at some
time) imposed restrictions on $U_{M}(t)$. Here the requirement of
sufficient strength becomes that of maximal strength which is the
same as completeness for this example. This quantum mechanical
model differs from a classical model in that the meaning domain
of the word states with meaning was limited to those strings
containing the word state. No word state $|\underline{W}\rangle$
was interpreted as implying any property of word string states not
containing $|\underline{W}\rangle.$ It was also possible for both
a word state with meaning and its negation to appear and not
violate consistency. In this case with multiple branching paths
of word string states, consistency merely required that no word
state with meaning and its negation state appear on the same path
(or word string state). It did not prevent them from appearing on
different paths. This option is not available in the classical
case with only one path present.

\subsection{\textbf{Meaning and Information}}
The emphasis here on the physical nature of language is to be contrasted with
Landauer's point that information is physical$^{(}$\cite{Landauer}$^{)}$. Landauer's
point stresses the physical nature of the information content of the states
representing expressions in a language.  Here the emphasis is directly on the physical
nature of language independent of the information content of any word string state or
formula string state.

The independence of the physical nature and information content of language is
emphasized by the observation that the relationship between the meaning, if any, of
language expression states and their algorithmic information content, is, at best,
complex and may be nonexistent. The proof, which follows that given
elsewhere$^{(}$\cite{BenUMLCQM}$^{)}$, consists of showing the unitary dynamics for
simulation of theorem enumerations for two theories with about the same algorithmic
complexity. But the theorems of one of the theories have meaning and those of the other
do not.

Let $U_{1}$ and $U_{2}$ represent the unitary dynamics for a
single time step (for simplicity a discrete time step model is
used here) of simulations of theorem enumeration machines for two
different theories, $T_{1}$ and $T_{2}$.  Such machines can be
modelled as described above by a complex head $M$ moving on a one
dimensional lattice of quantum systems, i.e. as a quantum Turing
machine.

There are different ways to describe what is meant by
simulation$^{(}$\cite{Bernstein,Berthiaume}$^{)}$.  However here it is sufficient to
require that for each $k=1,2$ the dispersion of the probability $|\langle
\underline{s(n)}|(U_{k})^{n}|\underline{\#},i,0\rangle |^{2}$ around the specific
expression state $|\underline{s_{k}(n)}\rangle$ is required to be less than $\epsilon$
for all $n<N\equiv N(\epsilon)$. The state $|\underline{s_{k}(n)}\rangle$ is a word
string state, Eq. \ref{wdstr}, in which each word state is a theorem state of $T_{k}$.

The $n$ dependence follows from the fact that theorem enumeration
systems described by the $U_{k}$ do not halt and, as $M$ moves
along the lattice, the number of theorem states in
$|\underline{s_{k}(n)}\rangle$ increases with increasing $n$. The
$n$ dependence in the states $|\underline{s(n)}\rangle$ shows that
as $n$ increases the length of the word string states which are
close to $|\underline{s_{k}(n)}\rangle$ also increases with $n$.
The dependence of $n$ on $\epsilon$ shows that the dispersion is
expected to increase with increasing $n$. This increase is
expressed by the condition that $N(\epsilon)$ gets smaller as
$\epsilon$ decreases.

Based on work in the literature$^{(}$\cite{Berthiaume,Vitanyi,Gacs}$^{)}$, the quantum
algorithmic complexity of the $U_{k}$ is described as the length of the shortest qubit
string states as input to a universal quantum Turing machine $U$ that give a simulation
of the $U_{k}$.  If $|\underline{q_{k}}\rangle$ are product qubit states such that
$U^{n}|\underline{q_{k}}\rangle$,acting on $|\underline{\#},i,0\rangle$, simulates
$(U_{k})^{n}$ acting on $|\underline{\#},i,0\rangle$ for $k=1,2$, then the algorithmic
complexity of $U_{k}$ is the length of the shortest state $|\underline{q_{k}}\rangle$
that gives the simulation. Again by simulation it is sufficient to require that the
dispersion of the probabilities $|\langle \underline{s(n)}|(U)^{n}
|\underline{q_{k}}\rangle|\underline{\#}i,0\rangle |^{2}$ around expression states
$|\underline{s_{k}(n)}\rangle$ for $k=1,2$ is less than $\epsilon$ for all $n<N\equiv
N(\epsilon)$ where the $|\underline{s_{k}(n)}\rangle$ is a theorem string state for
$T_{k}$.

This definition extends to quantum Turing machines the definition based on classical
machines$^{(}$\cite{Chaitin}$^{)}$ that defines the algorithmic complexity of a theory
as the length of  the shortest program as input to a universal machine that enumerates
the theorems of the theory. Since it is decidable whether a formula is or is not an
axiom, the states $|\underline{q_{k}}\rangle$ have finite lengths which are
proportional to the algorithmic complexities of the axiom sets $Ax_{1}$ and $Ax_{2}$
for the two theories$^{(}$\cite{Berthiaume}$^{)}$.

Let $Ax_{1}$ and $Ax_{2}$ be two sets of axioms that have about the same algorithmic
complexities and are such that the theory $T_{1}$ is consistent and $T_{2}$ is not
consistent.  It follows from G\"{o}del's completeness
theorem$^{(}$\cite{Godel,Shoenfield}$^{)}$ that, since $T_{1}$ is consistent, it has a
model in which the theorem states have meaning. Since $T_{2}$ is inconsistent, it has
no model so the theorem states enumerated by $U_{2}$ have no meaning. However the
algorithmic complexities of $U_{1}$ and $U_{2}$ are about the same since $Ax_{1}$ and
$Ax_{2}$ have about the same complexities.

The above proof depends on the existence of a theory $T_{2}$ that is inconsistent and
has about the same algorithmic complexity as $T_{1}$. To support this note that since
the number of sets of (nonlogical) axioms increases exponentially  with the algorithmic
complexity of the set in terms of the number of bits of the shortest program needed to
list the axioms$^{(}$\cite{Chaitin}$^{)}$, the number of theories increases
exponentially with the algorithmic complexity of the axiom sets. Also there are at
least as many inconsistent axioms sets as consistent ones. To see this let
$Ax^{\prime}_{1}$ be obtained from $Ax_{1}$ by adding the negation of a formula in
$Ax_{1}$ to the axioms in $Ax_{1}$.

$T^{\prime}_{1}$ is not a good candidate for $T_{2}$ because
algorithmic complexity of the proof of inconsistency for
$T^{\prime}_{1}$ is small. In particular the algorithmic
complexity of $T^{\prime}_{1}$ is bounded above by the length of
the shortest proof\footnote{A proof is a string of formulas such
that each formula in the string is either an axiom or is derived
from other formulas in the string by use of the logical rules of
deduction.} of inconsistency of $T^{\prime}_{1}$. An inconsistency
proof is a proof  that terminates with a formula that is the
negation of a formula appearing earlier in the string. It is clear
from this that the length of the shortest proof of inconsistency
for $T^{\prime}_{1}$ is quite small, and any formula enumeration
procedure is a theorem enumeration procedure.

$T_{2}$ must be a theory whose algorithmic complexity, in terms of
a decision procedure for the set $Ax_{2}$ of axioms, is about the
same as that of $T_{1}$ and is such that the length of the
shortest proof of inconsistency of $T_{2}$ is bounded below by the
algorithmic complexity of its own set of axioms.  This guarantees
that any theorem enumeration procedure  for $T_{2}$ has an
algorithmic complexity similar to that of $T_{1}$. This follows
from the observation that a suitable theorem enumeration procedure
that checks for consistency is such that it checks each terminal
formula in a proof to see if it is the negation of a formula
appearing earlier in the string. If not, the procedure continues.
If it is a negation, the procedure changes to a simple enumeration
procedure for all formulas.

additional justification for the existence of a theory $T_{2}$ with the requisite
properties is based on the observation  that the number of random sequences increases
exponentially with their length$^{(}$\cite{Chaitin}$^{)}$. It follows from this that
the number of proofs, even with the consistency check inserted as described above, that
are random should also increase exponentially with their length.  From this one
concludes that for large $n$ there should be many theories $T_{1}$ and $T_{2}$ with the
requisite properties.

\subsection{\textbf{G\"{o}del Maps}}\label{GM}
G\"{o}del maps have an interesting feature for a coherent theory,
or for any theory which is universally applicable to  physical
systems.  To see this it is worth a brief review of what a
G\"{o}del map is and what it does.

These maps are an important component of the proof of the G\"{o}del incompleteness
theorems for arithmetic$^{(}$\cite{Godel,Smullyan}$^{)}$.  In this case a G\"{o}del map
is a map $G$ from the expressions in the language of arithmetic to the natural numbers.
The map extends the domain of arithmetic in the sense that syntactic properties of
expressions in the language, which are not numbers, become properties of numbers. In
this way metamathematical properties of the theory become mathematical properties of
numbers which are in the domain of the theory. Then formulas of arithmetic can be
interpreted by anyone who knows $G$ as statements about properties of expressions in
the language of arithmetic.

Many different G\"{o}del maps are possible. An especially transparent one was first
described by Quine$^{(}$\cite{Quine}$^{)}$ (See also$^{(}$\cite{Smullyan}$^{)}$). The
map includes a one-one map $d:Sy\rightarrow {0,\cdots ,k-1}$ of the $k$ language
symbols in $Sy$ onto the $k$ digits of a $k-ary$ number representation. The map also
requires an association of lattice sites with powers of $k$. Typically one considers
expression states starting at site $0$. Then the site $j$ symbol state $|s,j\rangle$
corresponds to the number state $|d(s),j\rangle$ which corresponds to the number
$d(s)\times k^{j}$. If values of $j<0$ are allowed then the numbers correspond to
nonnegative $k-ary$ rational numbers with the $"k-al"$ point between $j=0$ and $j=-1$.

Such a map can be used to map expression states as tensor product states of the $13$
symbol states of the language described earlier into tensor product states of qukits
that correspond to a $k-ary$ representation of numbers.  Physically each symbol state
or qukit state often corresponds to states of single systems as in the spin projection
model described earlier.  However models can be considered in which each symbol state
in an expression product state correspond to an entangled state of several quantum
systems.\footnote{One can also construct representations of numbers in which there is
no correspondence between the tensor product representation of expression states or
qukit states and the states representing numbers. An example of this using complex
entangled states for $k=2$ was shown in$^{(}$\cite{BenEIPSRN}$^{)}$ for numbers $ <
2^{n}$ with $n$ arbitrary. A physical representation of numbers with entangled state
structure representing that in the example is very unlikely to exist.  The reason is
that a necessary condition that states of quantum systems represent numbers is that the
basic arithmetic operations be efficiently physically
implementable$^{(}$\cite{BenRNQM,BenRNQMALG,BenEIPSRN}$^{)}$.}  Examples of this are
shown by various quantum error correcting codes$^{(}$\cite{QEC}$^{)}$.

It is worth summarizing some of the main properties of G\"{o}del
maps $G$ in a more general context. For any mathematical theory
strong enough to include arithmetic, such as Zermelo Frankel set
theory, the maps $G$ extend the domain of the theory in the sense
that language expressions are mapped into the theory domain.  This
means that syntactic properties of expressions can be defined
through $G$, by properties of elements in the theory domain.

G\"{o}del maps can also be used in physical theories. However, for
these theories, they have some different properties.  For a
coherent theory of mathematics and physics, or for any physical
theory that is universally applicable, a G\"{o}del map does not
extend the domain of applicability of the theory.  The reason is
that, since language is physical, all expressions of any language
are already in the theory domain as states of physical systems.
For these theories a G\"{o}del map has a more limited function in
that it determines which of the many possible physical
representations of a language, if any, should be used to interpret
physical system states generated by some dynamical process as
expressions in some language.

Examples of these dynamical processes include creating written
text in a language, speaking, or, for microscopic systems,
generating quantum states by the action of some Hamiltonian that
can be interpreted  as expressions in some language.   However
other interpretations are possible and are often used. For example
the states could be interpreted as $k-ary$ representations of
numbers or be given some other interpretation. Which
interpretation is appropriate depends on the specific dynamics
being considered.

Macroscopic computers, which are in such wide use, are a good
illustration of these points. In these machines with a binary
code, strings of $0s$ and $1s$ can be stored and manipulated as
states of small magnetic regions located on some substrate.  It
makes no difference for the operation of the machine whether these
strings are considered as binary numbers or as expressions in a
language with a two letter alphabet.  This argument extends
immediately to $k-ary$ representations in that it makes no
difference as far as the dynamics of a particular machine is
concerned whether the small magnetic regions on the substrate
represent numbers or expressions in a language with a $k$ symbol
alphabet.

However, the usefulness of a  machine as a computer has everything
to do with the interpretation or meaning assigned to the input and
output states and with the details of the dynamical evolution of
the machine. Dynamical processes or machines that have useful
meaning for the number representations need not have a useful
meaning for the language expression representation. The converse
also holds in that dynamical processes or machines that have
meaning for language expressions have no useful meaning for
$k-ary$ representations of numbers.  An example of the former is a
machine that multiplies two numbers together to obtain the
product, all in a $k-ary$ representation. "Multiplying" two
language expressions together to obtain another is not a useful or
meaningful concept as far as syntactic properties of language
expressions are concerned. Conversely a dynamical process that
enumerates the proofs of theorems of a particular theory has no
useful meaning in number theory as an enumeration of a string of
extremely large numbers.

Interpretative maps play a similar role for microscopic quantum
mechanical machines or dynamical processes.  However, there is an
additional problem in that one must know what basis to use for the
interpretation. For example the spin projection model considered
earlier requires knowing the direction of the axis of quantization
in order to assign meaning to input and output states. Similarly,
to interpret states of quantum computers as numbers, one must know
the basis to use for the particular physical quantum system under
consideration.

This problem is especially important in the case of a complex
quantum system such as a quantum robot $M$ moving about in a
complex environment and interacting with various systems.  To
answer the question of whether $M$ is generating physical systems
in states with meaning and if so what the meaning is, requires not
only knowing which physical systems to examine but what basis
state representation to use. If one measures the states of
physical systems in the wrong basis then there is a good
probability that a misleading result will be obtained.  Also, as
is well known, the state of the measured system is changed so that
it cannot be reread to determine the original state.

As a specific example suppose $M$ is creating a product spin
projection state that, as a specific string of spin projections
along the $z$ axis, has meaning to him.  If an external observer
measures the state as a string of spin projections along the $x$
axis, then there is no way the observer can determine from his
results if the states generated by $M$ have meaning and, if so,
what the meaning is. The same holds for the results obtained by
$M$ for any reading of his own output after the measurements by
the external observer.
\\

\section{Discussion}\label{D}
At this point it is not known how to construct a coherent theory
of mathematics and physics.  However the material presented here
may help in that it should be regarded as a general framework for
constructing such a theory. The details and many aspects of a
coherent theory remain to be worked out. Also some or many of the
points and aspects described may need modification.  However some
of the points are expected to remain.

First and foremost among the remaining points is the requirement
that the coherent theory maximally describe its own validity and
sufficient strength and that it be valid and sufficiently strong.
This requirement is expected to greatly restrict the range of
allowed theories.  It may even be so restrictive that just one
theory satisfies it.

The requirement also has the advantage that it automatically
ensures that any theory satisfying it agrees with experiment. This
follows from the definition of validity, that any physical
property that is predictable by the theory and is testable by
experiment, is true.  Sufficient strength ensures that the theory
makes sufficiently many and powerful predictions to be recognized
as a coherent theory of physics and mathematics together.

This raises the problem that if the requirement that the theory
agree with experiment is built into the structure of the theory
itself, then one might think that the theory is not falsifiable or
even testable. This is not the case.  Even if the maximal validity
and sufficient strength requirement is built into the theory it
still must be tested. In particular the theory may be interpreted
to state, by complicated expressions, that it satisfies the
requirement. This would include a statement of  maximal agreement
with experiment. But is this in fact the case? Is the theory
statement of this true or false? One still has to carry out
experiments to find out.

One should also keep separate the requirement that the theory
maximally agree with experiment from what the actual results are
of carrying out the experiments.  For instance, incorporation of
the  validity and sufficient strength requirement into the theory
to the maximum extent possible may mean that the theory describes
the existence of a map between a set of theoretical predictions
and a set of experimental procedures, which are both described by
the theory. The theory would also describe general properties of
the map that correspond to agreement between theory and
experiment.

However existence and general description of such a map in a
coherent theory does not mean that a coherent theory is any
different than present day physics regarding the need to carry out
experiments to test the validity of theoretical predictions and
determine detailed properties of the map. A coherent theory that
maximally describes its own validity and and sufficient strength
may deepen the understanding between physics and mathematics, and
may even suggest new experiments, but it should not change the
status or need to carry out experiments. These are still needed to
see if the theory is valid and sufficiently strong.

It also may be the case that the most basic aspects of the
physical universe are a direct consequence of the basic
requirement that there exist a coherent theory that maximally
describes its own validity and completeness and is maximally valid
and complete. Included are the reasons why space-time is $3+1$
dimensional, why quantum mechanics is the correct physical theory,
and predictions of the existence and strengths of the four basic
forces.  However other aspects of the universe, which are also
predictable by the theory, are not in this category and are
subject to experimental test. This includes essentially all of the
experimental and theoretical work done in  physics.

There are also other possibilities to consider. For instance it
may be the case that, as discussed in subsection \ref{MDVSS},
there is no single coherent theory. Instead there may be a
nonterminating sequence of coherent theories, with each theory
more inclusive than those preceding it.  If this is the case then
the $n+1st$ theory may include in its domain the requirement that
the preceding $n$ theories all maximally agree with experiment.
But there may be other theoretical predictions in the $n+1st$
theory that are not present in the first $n$ theories whose
experimental status is outside the domain of the $n+1st$ theory.

Finally it should be noted that it may be worthwhile to replace validity in the basic
requirement with consistency.  In this case a coherent theory must maximally describe
its own consistency and sufficient strength and it must be consistent and sufficiently
strong. The advantage of this change is that, unlike the case for validity, consistency
can be defined purely syntactically by reference to proofs only. Also consistency is
related to semantic concepts through G\"{o}del's completeness
theorem$^{(}$\cite{Godel,Shoenfield}$^{)}$: a theory is consistent if and only if it
has a model. The usefulness of this alternate approach is a question for the future.
\\

\section*{Acknowledgements}
This work is supported by the U.S. Department of Energy, Nuclear
Physics Division, under contract W-31-109-ENG-38.

\appendix

\section{Appendix}

The physical representation of language considered here is based
on the presence or absence of systems in small potential wells
located on a two dimensional lattice of points on a solid state
matrix. The description is quite simple and will be limited to the
representation only. No dynamics corresponding to the generation
of word sequences, such as text or those that correspond to proofs
in a formal language, will be discussed.

The representation considered here is a model for text on printed
pages in that the systems in the potential wells are ink
molecules.  Each symbol corresponds to a specific pattern of
occupied wells surrounded by unoccupied wells. Expressions
correspond to paths of symbols on the lattice.   A solid state
matrix with all potential wells unoccupied corresponds to a blank
page. Locations of the wells on the page are given by $X,Y$
coordinates $x,y$. Multiple pages can be considered by extending
the lattice into three dimensions where $X-Y$ planes for different
values of $Z$ correspond to different pages.

Each potential well may or may not be occupied by ink molecules.
Here an ink molecule is a complex system with many closely spaced
internal states of excitation. The molecules are easily excited by
absorption of ambient light of all visible wavelengths, and the
excited states quickly decay by emitting cascades of infrared
photons as heat or by transfer of phonons to the solid state
matrix.

The state of an ink molecule in the ground state of the potential
well at $x,y$ in thermal equilibrium with an environment at
temperature $T$ is given by
\begin{equation}
\rho_{x,y} = \sum_{E}\frac{e^{-E/kT}}{Z}|E,0\rangle_{x,y}\langle
0,E|. \label{rhoxy} \end{equation}  Here $|E,0\rangle_{x,y}$
denotes the ink molecule in a state with excitation energy $E$ and
in the ground state of the well located at lattice site $x,y$. $Z$
is the partition function that normalizes the state. It is also
assumed that the combination of the shape and height of each
potential well and separation of the lattice points are such that
the states $|E,0\rangle_{x,y}$ and
$|E,0\rangle_{x^{\prime},y^{\prime}}$ are essentially orthogonal
whenever $x\neq x^{\prime}$ or $y\neq y^{\prime}$.

To keep things simple the assumption is made that the energy
spacing of the potential well states is large compared to $kT$
where $k$ is Boltzman's constant. Based on this Eq. \ref{rhoxy} is
a good approximation to the state of the ink molecule in a well at
$x,y$ as the probability of being in a state above the ground
state of the well is very small. It is also assumed that the
internal excitation state of an ink molecule is essentially
independent of whether the environment is visibly dark or well
illuminated with visible light, provided only that both
environments are at the same temperature.

The environmental bath also plays an important role in stabilizing
the position states of the individual ink molecules to eigenstates
of the individual potential wells.  For example ink molecule
states of the form
$$\sum_{E}\frac{e^{-E/kT}}{Z}\sum_{x,y,x^{\prime},y^{\prime}}
c_{x,y}c^{*}_{x^{\prime},y^{\prime}}|E,0\rangle_{x,y}\langle
0,E|_{x^{\prime},y^{\prime}}$$ would immediately decohere and
stabilize$^{(}$\cite{Zurek,Zeh}$^{)}$ to the diagonal form
$\sum_{x,y}|c_{x,y}|^{2}\rho_{x,y}$ with $\rho_{x,y}$ given by Eq. \ref{rhoxy}.

Let $\alpha$ be an arbitrary finite set of points on a lattice.
The quantum state corresponding to one ink molecule in each well
at all locations in $\alpha$ and all other wells unoccupied is
given by
\begin{equation}\rho_{\alpha} =  \bigotimes_{x,y\epsilon
\alpha}\rho_{x,y} \mbox{} =  \bigotimes_{x,y\epsilon
\alpha}\sum_{E_{x,y}}\frac{e^{-E_{x,y}/kT}}{Z}|E_{x,y},0\rangle_{x,y}\langle
0,E_{x,y}|. \label{rhoal}
\end{equation}

Symbols of a language correspond to sets of different patterns of
closely spaced occupied wells. To this end let $\alpha_{S}$ be the
set of occupied locations corresponding to the symbol $S$. A
potentially useful characterization of the set $\alpha_{S}$ is in
terms of a location $x,y$ that serves as a standard fiducial mark
or location parameter for the symbol, and a set $b$ of scaling and
other parameters needed to uniquely characterize the symbol $S$.
Using this notation, which replaces $\alpha_{S}$ by $S_{x,y,b}$,
Eq. \ref{rhoal} becomes
\begin{equation} \rho_{S_{x,y,b}}=\bigotimes_{x,y\epsilon
S_{x,y,b}}\rho_{x,y}. \label{rhoSxyb} \end{equation}

Some examples will serve to clarify this. The straight vertical
line extending for $n$ lattice sites in the $Y$ direction from
$x,y$ to $x,y+n-1$ corresponds to the symbol $"|"$ located at
$x,y$. The point $x,y$ locating one end of the symbol serves as a
fiducial location convention for this symbol. For each $x,y$ the
physical state of $"|"$ is given by $\rho_{|_{x,y,n}}=
\otimes_{x^\prime =x}^{x+n-1}\rho_{x^{\prime},y}$  Other examples
are the symbol $"/"$, a diagonal line of length $n$ whose state is
$\rho_{/_{x,y,n}}= \rho_{\alpha}$ with $\alpha
=\{x,y;x+1,y+1;\cdots ;x+n-1,y+n-1\}$, and the $"\top "$ symbol
with horizontal arm of length $2m+1$ and state description
$\rho_{\top_{x,y,n,m}}= \rho_{\alpha}$ where $\alpha =
\{x,y;\cdots ;x,y+n-1;x-m,y+n-1;\cdots ;x+m,y+n-1\}$. The values
of $n,m$ serve as scale factors for the symbols.  For example if
"$_{\top}$" is described by $n,m$, then "$\top$", which is the
same symbol but is twice as large, would be described by $2n,2m$.

These examples and Eq. \ref{rhoSxyb} show the physical state
representation for any printed symbol. This can be extended to
give the physical states of words which are strings of symbols. To
this end let $W(i)$ denote the $ith$ symbol of the word $W$
containing $N$ symbols. Let the function $\ell$ be a function from
the integers to locations on the lattice such that $\ell(i)$ is
the location of $W(i)$. Also $\alpha_{W(i),\ell(i)}$ denotes the
set of locations occupied by ink molecules for the symbol $W(i)$
at location $\ell(i)$. The sets $\alpha_{W(i),\ell(i)}$ are
disconnected sets where the (unoccupied) spacing between the sets
should be larger that the spacing, if any between the individual
ink molecule locations within each $\alpha_{W(i),\ell(i)}$.

The state $\rho_{W}$ for the word $W$ is given by
\begin{equation} \rho_{W} = \bigotimes_{i=1}^{N}\rho_{\alpha_{W(i),\ell(i)}}
 \end{equation}  where $\rho_{\alpha_{W(i),\ell(i)}}$ is given by Eq.
 \ref{rhoal}. In terms of fiducial marks and scale factors
\begin{equation}\rho_{W,\underline{x},\underline{y},\underline{b}}
=\bigotimes_{i=1}^{N}\rho_{W(i),\underline{x}(i)
,\underline{y}(i),\underline{b}(i)} \label{rhowd}\end{equation}
with $\rho_{W(i),\underline{x}(i)
,\underline{y}(i),\underline{b}(i)}$ given by Eq. \ref{rhoSxyb}.
Here $ \underline{x},\underline{y}, \underline{b}$ denote
functions from the symbols in $W$ to $x$ and $y$ lattice
positions, with $\ell(i) = \underline{x}(i),\underline{y}(i)$, and
to a set of scale factors for the symbols.

It is clear that this representation can be extended to strings of
words and to text in general.  In this case if $ \underline{W}$
denotes text or a string of $M$  words and $ \underline{\ell}$
denotes a lattice location function for the $M$ words in $
\underline{W}$  then \begin{equation} \rho_{ \underline{W},
\underline{\ell}} =\otimes_{j=1}^{M}\rho_{
\underline{W}(j),\underline{\ell}(j)}. \end{equation}

The location function $ \underline{\ell}$ determines how the words
are organized into text in the same way that the functions
$\ell(j)$ for each $j$ determine how the symbols in the $jth$ word
are organized into a word. For many languages, texts are often
organized into lines of symbols in one space direction  with
successive lines ordered in an orthogonal direction. Successive
pages are then ordered in the third space direction. Here spatial
distances between symbols, lines, and pages are used for the
ordering. This organization is reflected in the functions $
\underline{\ell}$ and $\ell(j)$.

The point of this  is to emphasize that, in this model, the functions
$\underline{\ell}$ and $\ell(j)$  are given by the rules used to read the words and
text. These rules are given by the dynamics of the reading process. The dynamics of
this process are not completely arbitrary.  They are subject to the requirement of
efficient physical implementation. This requirement, which was discussed elsewhere in
the context of representing numbers by states in quantum
mechanics$^{(}$\cite{BenRNQM,BenRNQMALG,BenEIPSRN}$^{)}$, means that there must be a
physical process which can read the text and that the space-time and thermodynamic
resources expended to implement the reading must be minimized. In particular the
resources expended must not be exponential in the number of symbols read.

It is worth discussing this in a bit more detail.   Consider for
example symbols scattered about on an infinite $X-Y$ lattice. Any
reading rule in which the determination of the  location for
reading the $n+1st$ symbol is based on what the first $n$ symbols
were requires an exponential amount of resources.   This is based
on the observation that if there are $m$ symbols in the language,
then for each $n$ the rule must distinguish among $m^{n}$
alternatives to make the determination.

Reading rules in use do not have this property in that
determination of the location of the $n+1st$ symbol from the value
of $n$ does not depend on the state of the first $n$ symbols.  As
such the rules are efficient in that the resources expended are
polynomial in the number of symbols read. Since the requirement of
polynomial efficiency is quite weak, there are many rules that
satisfy this condition, so one would want to pick rules that more
or less minimize the free energy resources expended. These are
the ones used in practice and include those used to read text on
a page.

The model described is clearly robust in the sense that each
reading of the text has a very small probability to change the
individual symbol states or move them about on the lattice.  Thus
it can be read many times with the cumulative probability for
changing the text remaining small. Physically this is a
consequence of the fact that photons in the visible light range
excite the ink molecules to internal excited states. The potential
wells and interactions with the component atoms in the molecules
are such that the amplitudes for exciting an ink molecule to an
excited well state, or to move it from one lattice location to
another, are very small. However, sufficiently many repeated
readings can move ink molecules around and make significant
changes in the quantum state of the text.


\begin{thebibliography}{99}

\bibitem{Weinberg}
S. Weinberg, {\it Dreams of a Final Theory}, (Vintage Books, New
York, NY, 1994).

\bibitem{Greene}
B. Greene, {\it The Elegant Universe}, (Vintage Books, New York,
NY 2000).

\bibitem{Casti}
J. Casti and A. Karlqvist, Editors, {\it Boundaries and Barriers,
On the Limits to Scientific Knowledge}, (Perseus Books, Reading,
MA, 1996).

\bibitem{Tegmark}
M. Tegmark, {\it Ann. Phys.} {\bf 270}, 1-51, (1998).

\bibitem{Shoenfield}
J.R. Shoenfield, {\it Mathematical Logic},  (Addison Weseley
Publishing Co. Inc., Reading, MA,  1967).

\bibitem{Smullyan}
R. Smullyan, {\it G\"{o}del's Incompleteness Theorems} (Oxford
University Press, New York, 1992).

\bibitem{Godel}
K. G\"{o}del, "Uber formal unentscheidbare S\"{a}tze der Principia
Mathematica und Vervandter Systeme I" {\it Monatschefte f\"{u}r
Mathematik und Physik}, {\bf 38}, 173-198, (1931).

\bibitem{Frankel}
A. A. Frankel, Y. Bar-hillel, A. Levy, and D. van Dalen, {\it
Foundations of Set Theory, second revised edition}, Studies in
Logic and the Foundations of Mathematics, Vol 67, (North-Holland
Publishing Co. Amsterdam, 1973).

\bibitem{BenRAN}
P. Benioff, {\it Jour. Math. Phys}. {\bf 17}, 618, 629, (1976).

\bibitem{Vaananen}
J. V\"{a}\"{a}n\"{a}nen, {\it Bull. Symbolic Logic}, {\bf 7}, pp
504-519, (2001).

\bibitem{BenDTVQM}
P. Benioff, {\it Phys. Rev. A} {\bf 59}, 4223 (1999).

\bibitem{Heyting}
A. Heyting, {\it Intuitionism, An Introduction, 3rd Revised
Edition} (North-Holland Publishing, New York, 1971).

\bibitem{Bishop}
E. Bishop, {\it Foundations of Constructive Analysis} (McGraw Hill
Book Co. New York, 1967).

\bibitem{Beeson}
M. J. Beeson {\it Foundations of Constructive Mathematics}
Metamathematical Studies, (Springer Verlag, New York, 1985).


\bibitem{Penrose}
R. Penrose, {\it The Emperor's New Mind}  (Penguin Books, New
York, 1991).

\bibitem{Hersh}
P. Davies and R. Hersh, {\it The Mathematical Experience},
(Birkh\"{a}user, Boston, 1981).

\bibitem{Kline}
M. Kline, {\it Mathematics, The Loss of Certainty}, (Oxford
University Press, New York, 1980).

\bibitem{MarMyc}
W.W Marek and J. Mycielski,  "The Foundations of Mathematics in
the Twentieth Century", {\it Amer. Math. Monthly}, {\bf 108} pp
449-468, (2001).

\bibitem{Papadimitriou}
C. H. Papadimitriou, {\it Computational Complexity} (Addison
Weseley Publishing Co. Reading MA 1994).

\bibitem{DeutschMLQ}
D.Deutsch, A. Ekert, and R. Lupacchini, {\it Bulletin Symbolic
Logic}, {bf 6}, 265-283 (2000).

\bibitem{Shor}
P. Shor, in {\it Proceedings, 35th Annual Symposium on the
Foundations of Computer Science}, S. Goldwasser (Ed), (IEEE
Computer Society Press, Los Alamitos, CA, 1994), pp 124-134.

\bibitem{Grover}
L. K. Grover, \textit{Phys. Rev. Letters}, {\bf 79} 325 (1997); G.
Brassard, \textit{Science} {\bf 275},627 (1997); L. K. Grover,
\textit{Phys. Rev. Letters}, {\bf 80} 4329 (1998).

\bibitem{Feynman}
R. P. Feynman, \textit{Int. Jour. Theoret. Phys}. {\bf 21} 467
(1982).

\bibitem{Zalka}
C. Zalka, "Efficient Simulation of Quantum Systems by Quantum
Computers", Los Alamos Archives preprint quant-ph/9603026.

\bibitem{Wiesner}
S. Wiesner, "Solutions of Many Body Quantum Systems by a Quantum
Computer", Los Alamos Archives preprint quant-ph/9603028.

\bibitem{Abrams}
D. S. Abrams and S. Lloyd, \textit{Phys. Rev. Letters}, {\bf 79},
pp. 2586-2589, (1997).

\bibitem{Lloyd}
S. Lloyd, \textit{Science}, {\bf 273}, pp. 1073-1078, (1996).

\bibitem{Landauer}
R. Landauer, \textit{Physics Today} {\bf 44}, No 5, 23, (1991);
\textit{Physics Letters A} {\bf 217} 188, (1996); in "Feynman and
Computation, Exploring the Limits of Computers", A.J.G.Hey, Ed.,
(Perseus Books, Reading MA, 1998).

\bibitem{Bridgman}
P. Bridgman, {\it Nature of Physical Theory}, (Dover Publications,
New York, 1936).

\bibitem{Svozil}
K. Svozil, \textit{Foundations of Physics }{\bf 25} 1541 (1995).

\bibitem{Finkelstein}
D. Finkelstein, Int. Jour. Theoret. Phys. {\bf 31} 1627 (1992);
"Quantum sets, assemblies and plexi" in \textit{Current Issues in
Quantum Logic}, E. Beltrametti and B. van Frassen, Eds, (Plenum
Publishing Co. New York, 1981), pp 323-333; {\it Quantum
Relativity} (Springer Verlag, New York, 1994).

\bibitem{Takeuti}
G. Takeuti, "Quantum Set theory" in \textit{Current Issues in
Quantum Logic}, E. G. Beltrametti and B. C. van Fraasen, Editors
(Plenum Press, New York 1981), pp 303-322.


\bibitem{Nishimura}
H. Nishimura, \textit{Int. Jour. Theoret. Phys.}, {\bf 32} 1293
(1993); {\bf 43} 229 (1995).

\bibitem{Odlyzko}
A. Odlyzko, "Primes, Quantum Chaos and Computers" in
\textit{Number Theory}, Proceeding of a Symposium, May 4 1989,
Washington D. C., Board on Mathematical Sciences, National
Research Council, 1990, pp 35-46.

\bibitem{Crandall}
R. E. Crandall, \textit{J. Phys. A: Math. Gen.} {\bf 29} 6795
(1996).

\bibitem{Okubo}
S. Okubo, "Lorentz-invariant Hamiltonian and Reimann Hypothesis"
Los Alamos Archives, quant-ph/9707036.

\bibitem{Ozhigov}
Y. Ozhigov, "Fast Quantum Verification for the Formulas of
Predicate Calculus", Los Alamos Archives Rept No. quant-
ph/9809015.

\bibitem{Buhrman}
H. Buhrman, R. Cleve, and A. Wigderson, "Quantum vs Classical
Communication and Computation", Los Alamos Archives Rept. No.
quant-ph/9802040.

\bibitem{Schmidhuber}
C. Schmidhuber, Los Alamos Archives preprint hep-th/0011065.

\bibitem{Blaha}
S. Blaha, "Cosmos and Consciousness", \textit{1stbooks Library,
Bloomington IN, 2000}; "A Quantum Computer Foundation for the
Standard Model and Superstring Theory", Los Alamos Archives Rept.
No. quant-ph/0201092.

\bibitem{Spector}
D. Spector, \textit{ Jour. Math. Phys.} {\bf 39}, 1919 (1998).

\bibitem{Foschini}
L. Foschini, "On the logic of Quantum Physics and the concept of
the time", Los Alamos Archives Preprint, quant-ph/9804040.

\bibitem{Mackey}
G. W. Mackey, "Mathematical Foundations of Quantum Mechanics", (W.
A Benjamin, Inc. New York, 1963).

\bibitem{Haag}
R. Haag, {\it Local Quantum Physics: fields, particles, algebras}
 (Springer-Verlag, Berlin, New York, 1992).

\bibitem{BirkhfVnN}
G. Birkhoff and J. von Neumann, \textit{Ann. Math.} {\bf 37}, 743
(1936).

\bibitem{Jauch}
J. M. Jauch and C. Piron, \textit{Helv. Phys. Acta}, {\bf 36}, 837
(1963); C. Piron, ibid. {\bf 37}, 439 (1964).

\bibitem{Hardy}
L. Hardy, Los Alamos Archives Preprint quant-ph/0101012.

\bibitem{Wigner}
E. Wigner, \textit{Commum. Pure and Applied Math.} {\bf 13} 001
(1960), Reprinted in E. Wigner, {\it Symmetries and Reflections},
(Indiana Univ. Press, Bloomington IN 1966), pp222-237.

\bibitem{Davies}
P. C. W. Davies, "Why is the Physical World so Comprehensible?" in
\textit{Complexity, Entropy, and Physical Information},
Proceedings of the 1988 workshop on complexity, entropy, and the
physics of information, may-june 1989, Santa Fe New Mexico, W. H.
Zurek, Editor, (Addison-Weseley Publishing Co. Redwood City CA
1990, pp 61-70).

\bibitem{Keisler}
C. C. Chang and H. J. Keisler, {\it Model Theory}, Studies in
Logic and the Foundations of Mathematics, Vol 73,  (American
Elsevier Publishing Co. Inc. New York, NY (1973)), PP. 36-45.

\bibitem{Chaitin}
G. Chaitin, {\it Information Theoretic Incompleteness}, World
Scientific Series in Computer Science, Vol. 35, (World Scientific
Publishing, Singapore, 1992); {\it Information Randomness \&
Incompleteness},  Series in Computer Science - Vol 8, Second
Edition, (World Scientific, Singapore, 1990); \textit{Scientific
American} {\bf 232} pp. 47-52, (1975); \textit{American
Scientist}, {\bf 90} pp 164-171, (2002).

\bibitem{Cohen}
P. J. Cohen, {\it Set Theory and the Continuum Hypothesis}, (W. A.
Benjamin, Inc. New York NY, (1966)).

\bibitem{Woodin}
H. Woodin, \textit{Notices, Amer. Math. Soc.}, {\bf 48}, pp
567-576, (2001).

\bibitem{Stillwell}
J. Stillwell, \textit{Amer. Math. Monthly}, {\bf 109}, pp.
286-298, (2002).

\bibitem{Bennettetal}
C.H. Bennett, D.P. DiVincenzo, T. Mor, P.W. Shor, J.A. Smolin,
B.M. Terhal, \textit{Phys.Rev.Lett.}, {\bf 82}, 5385 (1999); C. H.
Bennett, D. P. DiVincenzo, C. A. Fuchs, T. Mor, E. Rains, P. W.
Shor, J. A. Smolin, and W. K. Wootters, \textit{Phys. Rev A}, {\bf
59}, 1070 (1999).

\bibitem{BenEIPSRN}
P. Benioff,\textit{ Phys. Rev. A} to appear, November 2001; Los
Alamos archives quant-ph/0104061.

\bibitem{Turing}
A. Turing, \textit{Proc. London Math. Soc.} {\bf 42}, pp. 230-265,
(1936).

\bibitem{Deutsch}
D. Deutsch, \textit{Proc. Roy. Soc. London, Series A}, {\bf 400},
997 (1985).

\bibitem{Bernstein}
E. Bernstein and U. Vazirani, \textit{SIAM jour. Comput.}, {\bf
26}, pp. 1541-1557, (1997).

\bibitem{Tegmark1}
M. Tegmark, Los Alamos archives Preprint quant-ph/9709032.

\bibitem{Everett}
H. Everett, \textit{Reviews of Modern Physics}, {\bf 29},
454-462(1957).

\bibitem{Wheeler1}
J. A. Wheeler, \textit{Reviews of Modern Physics}, {\bf 29},
463-465, (1957).

\bibitem{Stapp}
H. P. Stapp, {\it Mind, Matter, and Quantum Mechanics}, (Springer
Verlag, Berlin 1993).

\bibitem{Squires}
E. Squires, {\it Conscious Mind in the Physical World} (IOP
Publishing, Bristol England, 1990).

\bibitem{Page}
D. Page,  Los Alamos Archives preprint quant-ph/0108039.

\bibitem{Wigner1}
E. Wigner, "Remarks on the Mind Body Question", in \textit{The
Scientist Speculates}, I. Good and W. Heinemann, Eds, (Putnam,
London, 1962).

\bibitem{Albert}
D. Albert, \textit{Physics Letters} {\bf 98A} 249 (1983);
\textit{Philosophy of science} {\bf 54} 577 (1987); "The Quantum
Mechanics of Self- measurement" in \textit{Complexity, Entropy and
the Physics of Information}, proceedings of the 1988 workshop in
Santa Fe, New Mexico, 1989, W. Zurek, Ed.  (Addison Wesely
Publishing Co. 1990).

\bibitem{BenRNQM}
P. Benioff, \textit{Phys. Rev. A} {\bf 63} \# 032305, (2001).

\bibitem{BenRNQMALG}
P. Benioff, Los Alamos Archives Preprint, quant-ph/0103078,
Accepted for publication in special issue of Algorithmica.

\bibitem{BarTip}
 J. D. Barrow and F. J. Tipler, {\it The Anthropic Cosmologic
Principle}, (Oxford University Press, 1989).

\bibitem{Hogan}
C. Hogan, \textit{Revs. Modern Phys}. {\bf 72}, 1149, (2000).

\bibitem{Greenstein}
G. Greenstein and A. Kropf, \textit{Am. Jour. Phys.} {\bf 58},
746, (1989).

\bibitem{BenUMLCQM}
P. Benioff, Los Alamos Archives preprint quant-ph/0106153.

\bibitem{BenQR}
P. Benioff, Phys. Rev. A, {\bf 58}, pp. 893-904, (1998).

\bibitem{Berthiaume}
A. Berthiaume, W. van Dam, and S. Laplante, \textit{Jour. Comput.
Syst. Sciences}, {\bf 63},pp. 201-221, (2001).

\bibitem{Vitanyi}
P. Vitanyi, "Three Approaches to the Quantitative Definition of
Information in an Individual Pure Quantum state", {\it Proc. 15th
IEEE Conference on Computational Complexity} Piscatawy NJ, (2000),
pp. 263-270; Los Alamos Archives preprint quant-ph/9907035.

\bibitem{Gacs}
P. Gacs, \textit{Jour. Phys. A, Math Gen.}, {\bf 34}, pp.
6859-6880, (2001).

\bibitem{Quine}
W. V.O. Quine, {\it Mathematical Logic}, (Norton, 1940).

\bibitem{QEC}
R. Laflamme, C. Miquel, J. P. Paz, and W. H. Zurek, \textit{Phys.
Rev. Letters}, {\bf 77} 198 (1996); D. P. DiVincenzo and P. W.
Shor, \textit{Phys. Rev. Letters}, {\bf 77}3260 (1996); E. M.
Raines, R. H. Hardin, P. W. Shor, and N. J. A. Sloane,
\textit{Phys. Rev. Letters} {\bf 79} 954 (1997).

\bibitem{Zurek}
W. H. Zurek, \textit{Phys. Rev. D }{\bf 24} 1516, (1981); {\bf 26}
1862 (1982).

\bibitem{Zeh}
E. Joos and H. D. Zeh, \textit{Zeit. Phys. B} {\bf 59}, 23,
(1985); H. D. Zeh quant-ph/9905004; E Joos, quant-ph/ 9808008.

\end{thebibliography}
\end{document}